\documentclass[thmsa,12pt]{article}
\usepackage{amsfonts}

\usepackage{amsmath}

\input{tcilatex}

\begin{document}

\setcounter{page}{0} \topmargin0pt \oddsidemargin5mm \renewcommand{%
\thefootnote}{\fnsymbol{footnote}} \newpage \setcounter{page}{0} 
\begin{titlepage}
\begin{flushright}
EMPG-04-02\\
\end{flushright}
\vspace{0.5cm}
\begin{center}
{\Large {\bf Auxiliary matrices for the six-vertex model\\ and the algebraic Bethe ansatz } }

\vspace{0.8cm}
{ \large Christian Korff}

\vspace{0.5cm}
{\em School of Mathematics, University of Edinburgh\\
Mayfield Road, Edinburgh EH9 3JZ, UK}
\end{center}
\vspace{0.2cm}
 
\renewcommand{\thefootnote}{\arabic{footnote}}
\setcounter{footnote}{0}

\begin{abstract}
We connect two alternative concepts of solving integrable models, Baxter's 
method of auxiliary matrices (or $Q$-operators) and the algebraic Bethe ansatz. The main 
steps of the calculation are performed in a general setting and a formula for the Bethe 
eigenvalues of the $Q$-operator is derived. A proof is given for states which contain up 
to three Bethe roots. Further evidence is provided by relating the findings to the six-vertex 
fusion hierarchy. For the  XXZ spin-chain we analyze the cases when the deformation 
parameter of the underlying quantum group is evaluated both at and away from a root of unity.
\end{abstract}
\vfill{ \hspace*{-9mm}
\begin{tabular}{l}
\rule{6 cm}{0.05 mm}\\
c.korff@ed.ac.uk
\end{tabular}}
\end{titlepage}
\newpage 

\section{Introduction}

This paper is a continuation of two previous works \cite{KQ,KQ2} on the
six-vertex model and the associated XXZ Heisenberg spin-chain at roots of
unity. That is, we consider the integrable model defined via the Hamiltonian 
\begin{equation}
H=\sum_{m=1}^{M}\sigma _{m}^{+}\sigma _{m+1}^{-}+\sigma _{m}^{-}\sigma
_{m+1}^{+}+\frac{q+q^{-1}}{4}\,\sigma _{m}^{z}\sigma _{m+1}^{z},\quad \sigma
_{M+1}^{\pm }\equiv \lambda ^{\pm 2}\sigma _{1}^{\pm },\;\sigma
_{M+1}^{z}\equiv \sigma _{1}^{z}\;.  \label{Ham}
\end{equation}
Here $\{\sigma ^{x},\sigma ^{y},\sigma ^{z}\}$ are the Pauli matrices with $%
\sigma ^{\pm }=(\sigma ^{x}\pm i\sigma ^{y})/2$ and $\lambda \in \mathbb{C}$
fixes the boundary conditions. The anisotropy parameter in front of the
third term in (\ref{Ham}) is fixed in terms of the complex variable $q$
which in general will be of modulus one. Of particular interest to our
discussion is the case when $q$ is a primitive root of unity, i.e. $q^{N}=1$
for some integer $N>2$. ( We exclude here the cases $N=1,2$ which are
related to the XXX model.) At these particular values the above Hamiltonian
as well as the associated six-vertex transfer matrix exhibit extra
degeneracies in their spectra which are linked to an underlying
infinite-dimensional non-abelian symmetry algebra. At periodic boundary
conditions, $\lambda =1$, and in the commensurate sectors $2S^{z}=0\func{mod}%
N$ this symmetry algebra is isomorphic to the loop algebra of $sl_{2}$. This
has been first established by Deguchi, Fabricius and McCoy in \cite{DFM},
where additional results for $N=3,4$ outside the commensurate sectors can be
found. When $\lambda \neq 1$ the symmetry algebra will in general reduce to
the upper or lower Borel subalgebra \cite{D04,Ktw}. A special case is
obtained when the boundary conditions are tuned to $\lambda =q^{\pm S^{z}}$.
Then the symmetry algebra and the explicit form of its generators are known
for all spin sectors and integers $N$ \cite{Ktw}.\smallskip 

The presence of an non-abelian symmetry algebra is of interest as it now
allows one to connect the Bethe ansatz and representation theory. Note that
while this idea is not new, the mentioned symmetries are
infinite-dimensional and exist at finite length of the chain, $M<\infty $.
The combination of these two distinct features distinguishes the present
discussion from previously considered cases in the literature at open
boundary conditions \cite{PS} or at infinite volume \cite{JM}.\smallskip

An important first step in connecting the Bethe ansatz with the
representation theory of the aforementioned symmetry algebras is to find an
efficient way to analyze the structure of the degenerate eigenspaces of the
Hamiltonian as well as the transfer matrix. This can be achieved by using
Baxter's concept of auxiliary matrices \cite{Bx71}, also known as $Q$%
-operators, which satisfy certain operator functional equations with the
transfer matrix. The concept of auxiliary matrices has primarily received
attention in the context with the eight-vertex model \cite
{Bx72,Bx73a,Bx73b,Bx73c}; see \cite{FM8v,FM8v2} for a recent discussion
corresponding to the root of unity case.

In connection with the six-vertex model the subject has obtained fresh
impetus from new methods of constructing such $Q$-operators which provide an
alternative to Baxter's procedure described in e.g. \cite{BxBook}. These new
methods are based on the representation theory of quantum groups \cite
{BLZ97,BLZ99,AF97,RW02,KQ}. The latter method will be of importance to us as
the auxiliary matrices constructed in this manner, see \cite{KQ,KQ2}, have
been shown for several examples to yield information on the irreducible
representations of the symmetry algebra at roots of unity. Several facts
about the spectrum of the auxiliary matrices in \cite{KQ2} have so far only
been rigorously proven for $N=3$ and conjectured to hold true for $N>3$
employing numerical calculations.

The purpose of this article is to lend further support to the earlier
conjectures and to extend the discussion from auxiliary matrices with
periodic boundary conditions ($\lambda =1$) to quasi-periodic ones ($\lambda
\neq 1$) in order to accommodate the findings in \cite{Ktw}. To this end we
take a broader point of view and consider the connection between the
algebraic Bethe ansatz \cite{QISM} and auxiliary matrices in a general
setting and at generic values of $q$.

Away from a root of unity this will enable us to calculate the spectrum of
the auxiliary matrices constructed in \cite{RW02} and resolve certain
convergence problems due to an infinite-dimensional auxiliary space. We will
also make contact with the discussion in \cite{BLZ99}.

At roots of unity \cite{KQ,KQ2} the Bethe ansatz analysis will not yield the
complete set of eigenvalues for the $Q$-operators, as at most the highest
weight state in each degenerate eigenspace of the transfer matrix ought to
be a proper Bethe state, i.e. a state parameterized by finite solutions of
the Bethe ansatz equations. Detailed explanations will be given in the text.
In addition to the comparison with the algebraic Bethe ansatz we will also
make contact with the fusion hierarchy of the six-vertex model. The latter
provides an infinite series of higher-spin transfer matrices which can be
successively generated through a functional equation. At roots of unity this
series truncates and we will show how the auxiliary matrices are related to
the fusion matrices, similar to the discussion in \cite{FM8v2} for the
eight-vertex model. This will provide additional evidence for the spectrum
of the auxiliary matrices constructed in \cite{KQ,KQ2}.

The paper is organized as follows. In Section 2 we introduce the monodromy
matrices of the six-vertex transfer matrix and the $Q$-operators constructed
in \cite{RW02} and \cite{KQ,KQ2}. Afterwards we derive the commutation
relations between the auxiliary matrices and the Yang-Baxter algebra.

This sets the stage to compute the action of the auxiliary matrices on Bethe
states in Section 3. We will find that they are eigenstates of the auxiliary
matrices and compute the corresponding eigenvalues. This is done without
specifying the quantum space. The computations are quite lengthy and
cumbersome whence we will only give a proof for Bethe states which contain
up to three Bethe roots. For the general case of an arbitrary number of
Bethe roots we formulate a conjecture.

In Section 4 we specialize our findings to the XXZ spin-chain. While we
discuss both $q$ being a root of unity and $q$ not being a root of unity,
the former case is considered in more detail in light of the aforementioned
symmetries. We check the conjectured formula for the eigenvalues of the $Q$%
-operators for consistency by inserting them into the respective functional
equation with the transfer matrix.

In Section 5 we present further support for the conjecture regarding the
spectrum of the auxiliary matrices by showing that the eigenvalues are also
consistent with the fusion hierarchy of the six-vertex model. The fusion
hierarchy is solved explicitly in terms of Bethe roots and we derive its
``truncation'' at roots of unity.

Section 6 contains the conclusions.

\section{Monodromy matrices}

The first step in the application of the algebraic Bethe ansatz is the
definition of the monodromy matrices. The monodromy matrices act on a tensor
product of two spaces, $\mathcal{H}_{0}\otimes \mathcal{H}$; the first is
called the auxiliary apce and the second the quantum space. Depending on the
choice of the auxiliary space we obtain the monodromy matrices for the $Q$%
-operator and the six-vertex transfer matrix. Since the monodromy matrices
are subject to the Yang-Baxter equation they can be constructed in the
framework of quantum groups. Let \{$e_{i},f_{i},q^{h_{i}}$\}$_{i=0,1}$ be
the Chevalley-Serre generators of the quantum loop algebra $U_{q}(\widetilde{%
sl}_{2})$ subject to the relations 
\begin{equation}
q^{h_{i}}e_{j}q^{-h_{i}}=q^{\mathcal{A}_{ij}}e_{j},\quad
q^{h_{i}}f_{j}q^{-h_{i}}=q^{-\mathcal{A}_{ij}}f_{j},\quad
q^{h_{i}}q^{h_{j}}=q^{h_{j}}q^{h_{i}},\quad i,j=0,1,  \label{AQG}
\end{equation}
where the Cartan matrix reads 
\begin{equation*}
\mathcal{A}=\left( 
\begin{array}{cc}
2 & -2 \\ 
-2 & 2
\end{array}
\right) \;.
\end{equation*}
We will only be dealing with representations where the central charge of the
affine extension is zero, whence we set 
\begin{equation}
h\equiv h_{1}=-h_{0}\;.
\end{equation}
In addition, for $i\neq j$ the Chevalley-Serre relations hold, 
\begin{eqnarray}
e_{i}^{3}e_{j}-[3]_{q}e_{i}^{2}e_{j}e_{i}+[3]_{q}e_{i}e_{j}e_{i}^{2}-e_{j}e_{i}^{3} &=&0,
\notag \\
f_{i}^{3}f_{j}-[3]_{q}f_{i}^{2}f_{j}f_{i}+[3]_{q}f_{i}f_{j}f_{i}^{2}-f_{j}f_{i}^{3} &=&0\;.
\label{CS}
\end{eqnarray}
The quantum algebra can be made into a Hopf algebra, its most important
property being that of a coproduct which we choose to be ($i=0,1$) 
\begin{equation}
\Delta (e_{i})=1\otimes e_{i}+q^{h_{i}}\otimes e_{i},\quad \Delta
(f_{i})=f_{i}\otimes q^{-h_{i}}+1\otimes f_{i},\quad \Delta
(q^{h_{i}})=q^{h_{i}}\otimes q^{h_{i}}\;.  \label{cop}
\end{equation}
The opposite coproduct $\Delta ^{\text{op}}$ is obtained by permuting the
two factors. If the deformation parameter $q$ is considered to be an
abstract indeterminate there exists the universal $R$-matrix intertwining
these two coproduct structures 
\begin{equation}
\mathbf{R\,}\Delta (x)=\Delta ^{\text{op}}(x)\,\mathbf{R},\quad x\in U_{q}(%
\widetilde{sl}_{2}),\quad \mathbf{R}\in U_{q}(b_{+})\otimes U_{q}(b_{-})\;.
\label{inter}
\end{equation}
Here $U_{q}(b_{\pm })$ denote the upper and lower Borel subalgebra,
respectively. We now define the six-vertex monodromy matrix by setting 
\begin{equation}
\mathbf{T}(z)=(\pi _{z}^{(1)}\otimes \pi _{\mathcal{H}})\lambda ^{h\otimes 1}%
\mathbf{R}\in \limfunc{End}\mathbb{C}^{2}\otimes \mathcal{H},  \label{mom6v}
\end{equation}
with $\pi _{z}^{(1)}:U_{q}(\widetilde{sl}_{2})\rightarrow \limfunc{End}%
\mathbb{C}^{2}$ being the fundamental evaluation representation given in
terms of Pauli matrices, 
\begin{eqnarray}
\pi _{z}^{(1)}(e_{0}) &=&z\sigma ^{-},\quad \;\pi
_{z}^{(1)}(f_{0})=z^{-1}\sigma ^{+},\quad \;\pi
_{z}^{(1)}(q^{h_{0}})=q^{-\sigma ^{z}},  \notag \\
\pi _{z}^{(1)}(e_{1}) &=&\sigma ^{+},\quad \;\pi _{z}^{(1)}(f_{1})=\sigma
^{-},\quad \;\pi _{z}^{(1)}(q^{h_{1}})=q^{\sigma ^{z}}\;.  \label{pi0}
\end{eqnarray}
We will denote the basis in the corresponding representation space $\mathbb{C%
}^{2}$ by $\{\left| 0\right\rangle ,\left| 1\right\rangle \}$. The
representation $\pi _{\mathcal{H}}$ in the second factor determines the
quantum space of our theory. At the moment we leave it unspecified in order
to emphasize the general nature of the following discussion. Later we will
set $\mathcal{H}=\left( \mathbb{C}^{2}\right) ^{\otimes M}$ and $\pi _{%
\mathcal{H}}=\bigotimes_{m=1}^{M}\mathbb{\pi }_{\zeta _{m}}^{(1)}$ in order
to obtain the inhomogeneous XXZ spin-chain. The factor in front of the
universal $R$-matrix involving $\lambda $ fixes the boundary conditions.

In the context of the algebraic Bethe ansatz it is customary to decompose
the monodromy matrix w.r.t. the auxiliary space $\mathbb{C}^{2}$, i.e. one
introduces the following elements of $\limfunc{End}\mathcal{H}$, 
\begin{equation}
A=\left\langle 0|\mathbf{T}|0\right\rangle _{\mathbb{C}^{2}},\quad
B=\left\langle 0|\mathbf{T}|1\right\rangle _{\mathbb{C}^{2}},\quad
C=\left\langle 1|\mathbf{T}|0\right\rangle _{\mathbb{C}^{2}},\quad
D=\left\langle 1|\mathbf{T}|1\right\rangle _{\mathbb{C}^{2}}\;.  \label{ABCD}
\end{equation}
Here the subscript $\mathbb{C}^{2}$ indicates that the matrix elements are
taken w.r.t. the first factor in (\ref{mom6v}). These elements obey the
familiar commutation relations of the six-vertex Yang-Baxter algebra which
are deduced from the relation 
\begin{equation}
R_{12}(w/z)\mathbf{T}_{1}(w)\mathbf{T}_{2}(z)=\mathbf{T}_{2}(z)\mathbf{T}%
_{1}(w)R_{12}(w/z),  \label{RTT}
\end{equation}
with 
\begin{equation}
R(z,q)=\tfrac{a+b}{2}1\otimes 1+\tfrac{a-b}{2}\,\sigma ^{z}\otimes \sigma
^{z}+c\,\sigma ^{+}\otimes \sigma ^{-}+c^{\prime }\sigma ^{-}\otimes \sigma
^{+}  \label{R}
\end{equation}
denoting the six-vertex $R$-matrix. The parameterization of the Boltzmann
weights is chosen in accordance with (\ref{cop}) and (\ref{pi0}), 
\begin{equation}
a=1,\quad b=\frac{1-z}{1-zq^{2}}\,q,\quad c=\frac{1-q^{2}}{1-zq^{2}},\quad
c^{\prime }=c\,z\;.  \label{abc}
\end{equation}

We next define the monodromy matrix from which we will obtain below the
auxiliary matrix or $Q$-operator. We follow an analogous procedure to that
just given with the difference that we now change to a higher dimensional
auxiliary space $\mathcal{H}_{0}$ belonging to a suitably chosen
representation $\pi _{w}^{\prime }:U_{q}(b_{+})\rightarrow \limfunc{End}%
\mathcal{H}_{0}$. The auxiliary space $\mathcal{H}_{0}$ will turn out to be
finite-dimensional at $q^{N}=1$ and infinite-dimensional when $q$ is not a
root of unity. The monodromy matrix is now 
\begin{equation}
\mathbf{Q}(w)=(\pi _{w}^{\prime }\otimes \pi _{\mathcal{H}})\lambda
^{h\otimes \mathbf{1}}\mathbf{R}\in \limfunc{End}\mathcal{H}_{0}\otimes 
\mathcal{H}\;\mathbf{.}  \label{momQ}
\end{equation}

At roots of unity there arises a technical subtlety. The universal R-matrix
need not exist as whether one can construct an intertwiner depends on the
precise nature of the representations $\pi _{w}^{\prime }$,$\pi _{\mathcal{H}%
}$. This is due to the enlarged centre of the quantum group at roots of
unity, see e.g. \cite{CPbook}. However, we will assume that the
representations are always chosen such that this is the case. Then the
definition (\ref{momQ}) has to be replaced by this intertwiner which is to
be explicitly constructed. (In the case of the monodromy matrix (\ref{mom6v}%
) one can first evaluate it away from a root of unity and then safely take
the root of unity limit.)

Similar to the six-vertex monodromy matrix we may also decompose (\ref{momQ}%
) over the auxiliary space, 
\begin{equation}
\mathbf{Q}=(Q_{ij})\quad \text{with}\quad Q_{ij}:=\left\langle i|\mathbf{Q}%
|j\right\rangle _{\mathcal{H}_{0}}\in \limfunc{End}\mathcal{H},
\label{Qmatrix}
\end{equation}
where the states are labelled by integers lying in a finite interval, $0\leq
i,j\leq N^{\prime }-1$, at roots of unity $q^{N^{\prime }}=\pm 1$ or in an
infinite interval $i,j\in \mathbb{Z}$ when $q^{N^{\prime }}\neq \pm 1$. Here
and in the following we set $N^{\prime }=N$ if the order of the root of
unity is odd and $N^{\prime }=N/2$ when it is even.

We will specify the respective auxiliary spaces momentarily. Before so doing
we introduce a third operator $L\in \limfunc{End}\mathcal{H}_{0}\otimes 
\mathbb{C}^{2}$ which intertwines the tensor product $\pi _{w}^{\prime
}\otimes \pi _{z}^{(1)}$ and satisfies together with the monodromy matrices (%
\ref{mom6v}) and (\ref{momQ}) the Yang-Baxter equation, i.e. 
\begin{equation}
L_{12}(w/z)\mathbf{Q}_{1}(w)\mathbf{T}_{2}(z)=\mathbf{T}_{2}(z)\mathbf{Q}%
_{1}(w)L_{12}(w/z)\;.  \label{YBE}
\end{equation}
Here we have assumed that we are only allowing for representations $\pi
_{w}^{\prime }$ in (\ref{momQ}) for which 
\begin{equation}
\lbrack L(w),\pi _{w}^{\prime }(h)\otimes 1+1\otimes \sigma ^{z}]=0\;.
\end{equation}
This restriction is necessary in order to accommodate the quasi-periodic
boundary conditions when $\lambda \neq 1$. Again, we decompose the $L$%
-operator, this time over the second factor, the two-dimensional space
associated with the evaluation representation $\pi _{z}^{(1)}$, 
\begin{equation}
L=\left( 
\begin{array}{cc}
\mathbf{\alpha } & \mathbf{\beta } \\ 
\mathbf{\gamma } & \mathbf{\delta }
\end{array}
\right) ,\quad \mathbf{\alpha },\mathbf{\beta },\mathbf{\gamma },\mathbf{%
\delta }\in \limfunc{End}\mathcal{H}_{0}\;.  \label{L}
\end{equation}
We shall now explicitly specify the evaluation representations $\pi
_{w}^{\prime }$ at and away from a root of unity as well as the matrix
elements of the $L$-operator within these representations.

\subsection{The auxiliary space when $q^{N}=1$}

We adopt the conventions in \cite{KQ2} and define the following evaluation
representation $\pi _{w}^{\prime }\equiv \pi _{w}^{\mu }$ of the full affine
quantum algebra $U_{q}(\widetilde{sl}_{2})$ at $q^{N}=1$ or equivalently $%
q^{N^{\prime }}=\pm 1$. Let $0\leq n\leq N^{\prime }-1,\;\mu \in \mathbb{C}$
and set 
\begin{eqnarray}
\pi _{w}^{\mu }(f_{1})\left| n\right\rangle &=&\left| n+1\right\rangle
,\quad \pi _{w}^{\mu }(e_{0})=w\,\pi _{w}^{\mu }(f_{1}),\quad  \notag \\
\pi _{w}^{\mu }(e_{1})\left| n\right\rangle &=&\frac{\mu +\mu ^{-1}-\mu
q^{2n}-\mu ^{-1}q^{-2n}}{(q-q^{-1})^{2}}\left| n-1\right\rangle ,\quad \pi
_{w}^{\mu }(f_{0})=w^{-1}\,\pi _{w}^{\mu }(e_{1}),  \notag \\
\pi _{w}^{\mu }(q^{h_{1}})\left| n\right\rangle &=&\mu ^{-1}q^{-2n-1}\left|
n\right\rangle ,\quad \pi _{w}^{\mu }(q^{h_{0}})=\pi _{w}^{\mu
}(q^{-h_{1}})\;.  \label{pi1}
\end{eqnarray}
The non-vanishing matrix elements of the $L$-operator then read \cite{KQ} 
\begin{eqnarray}
\alpha _{n} &=&\left\langle n|\mathbf{\alpha }|n\right\rangle =(w/z)\,\mu ^{-%
\frac{1}{2}}q^{-n+\frac{1}{2}}-\mu ^{\frac{1}{2}}q^{n+\frac{1}{2}},  \notag
\\
\delta _{n} &=&\left\langle n|\mathbf{\delta }|n\right\rangle =(w/z)\,\mu ^{%
\frac{1}{2}}q^{n+\frac{3}{2}}-\mu ^{-\frac{1}{2}}q^{-n-\frac{1}{2}},  \notag
\\
\gamma _{n} &=&\left\langle n|\mathbf{\gamma }|n+1\right\rangle =\mu ^{\frac{%
1}{2}}q^{n+\frac{3}{2}}\frac{\mu +\mu ^{-1}-\mu q^{2n+2}-\mu ^{-1}q^{-2n-2}}{%
q-q^{-1}},\quad n<N^{\prime }-1,  \notag \\
\beta _{n} &=&\left\langle n|\mathbf{\beta }|n-1\right\rangle
=(w/z)\,(q-q^{-1})\mu ^{-\frac{1}{2}}q^{-n+\frac{1}{2}},\quad n>0\;.
\label{G1}
\end{eqnarray}
Note that this representation can be extended to generic $q$ but is then
infinite-dimensional, i.e. $n\in \mathbb{Z}_{\geq 0}$. The reason for this
particular choice of the representation is explained in \cite{KQ,KQ2}. Here
we simply recall that the following exact sequence holds \cite{KQ}, 
\begin{equation}
0\rightarrow \pi _{w^{\prime }}^{\mu q}\overset{\imath }{\hookrightarrow }%
\pi _{w}^{\mu }\otimes \pi _{z}^{(1)}\overset{\tau }{\rightarrow }\pi
_{w^{\prime \prime }}^{\mu q^{-1}}\rightarrow 0,\quad w=w^{\prime
}q^{-1}=w^{\prime \prime }q=z/\mu ,  \label{seq1}
\end{equation}
with the inclusion $\imath $\ and the projection $\tau $ detailed in \cite
{KQ} (cf Section 4.1 and 4.2). From this decomposition of the tensor product
one now derives a functional equation of the following type 
\begin{equation}
T(z)Q_{\mu }(w)=\phi _{1}(z)\,Q_{\mu q}(w^{\prime })+\phi _{2}(z)\,Q_{\mu
q^{-1}}(w^{\prime \prime })  \label{TQ1}
\end{equation}
where 
\begin{equation}
T=(\limfunc{Tr}_{\mathbb{\pi }_{z}^{(1)}}\otimes 1_{\mathcal{H}})\mathbf{T}%
=A+D\mathbf{\quad }\text{and\quad }Q_{\mu }=(\limfunc{Tr}_{\mathbb{\pi }%
_{w}^{\mu }}\otimes 1_{\mathcal{H}})\mathbf{Q}=\sum_{n=0}^{N^{\prime
}}Q_{nn}\;.  \label{T&Q}
\end{equation}
Here $\phi _{1},\phi _{2}$ are some coefficient functions whose precise form
depends on the choice of the quantum space $\mathcal{H}$ which is as yet
unspecified. For the XXZ spin-chain we will present them below.

\subsection{Infinite-dimensional auxiliary space for $q^{N}\neq 1$}

Rossi and Weston introduced in \cite{RW02} the following
infinite-dimensional four-parameter representation $\pi ^{+}=\pi
^{+}(w;s_{0},s_{1},s_{2})$ of the upper Borel subalgebra $U_{q}(b_{+})$, 
\begin{eqnarray}
\pi ^{+}(e_{1})\left| n\right\rangle &=&\left| n-1\right\rangle ,\quad \pi
^{+}(q^{h_{1}})\left| n\right\rangle =s_{0}q^{-2n}\left| n\right\rangle
,\quad \pi ^{+}(q^{h_{0}})=\pi ^{+}(q^{-h_{1}}),  \notag \\
\pi ^{+}(e_{0})\left| n\right\rangle &=&\left( s_{1}s_{2}\frac{(q-q^{-1})^{2}%
}{w}+\frac{w}{(q-q^{-1})^{2}}+s_{1}q^{2n}+s_{2}q^{-2n}\right) \left|
n+1\right\rangle \;.  \label{pi2}
\end{eqnarray}
Here $n\in \mathbb{Z}$ and the representation space is thus
infinite-dimensional. For convenience we rescale the parameters according to 
\begin{equation}
s_{1,2}\rightarrow r_{1,2}=-(q-q^{-1})^{2}w^{-1}s_{1,2}\,,\;\quad \quad
s_{0}\rightarrow r_{0}=s_{0}\;.
\end{equation}
The matrix elements for the intertwiner $L$ are now calculated to be \cite
{RW02} 
\begin{eqnarray}
\alpha _{n} &=&\left\langle n|\mathbf{\alpha }|n\right\rangle
=(w/z)\,r_{2}r_{0}^{-\frac{1}{2}}q^{-n+2}-r_{0}^{-\frac{1}{2}}q^{n},  \notag
\\
\delta _{n} &=&\left\langle n|\mathbf{\delta }|n\right\rangle
=(w/z)\,r_{1}r_{0}^{\frac{1}{2}}q^{n}-r_{0}^{\frac{1}{2}}q^{-n},  \notag \\
\gamma _{n} &=&\left\langle n|\mathbf{\gamma }|n+1\right\rangle
=(q-q^{-1})r_{0}^{-\frac{1}{2}}q^{n+1},  \notag \\
\beta _{n} &=&\left\langle n|\mathbf{\beta }|n-1\right\rangle =(w/z)\,r_{0}^{%
\frac{1}{2}}q^{-n+1}\frac{r_{1}r_{2}+1-r_{1}q^{2n-2}-r_{2}q^{-2n+2}}{q-q^{-1}%
}\;.  \label{G2}
\end{eqnarray}
Similar to the root of unity case one has a decomposition of the tensor
product $\pi ^{+}\otimes \pi _{z}^{(1)}$ according to the exact sequence 
\cite{RW02} 
\begin{equation}
0\rightarrow \pi ^{+}(wq^{2};\mathbf{r}^{+})\overset{\imath }{%
\hookrightarrow }\pi ^{+}(w;\mathbf{r})\otimes \pi _{z}^{(1)}\overset{\tau }{%
\rightarrow }\pi ^{+}(wq^{-2};\mathbf{r}^{-})\rightarrow 0,\quad w=z,
\label{seq2}
\end{equation}
with 
\begin{equation}
\mathbf{r}=(r_{0},r_{1},r_{2})\quad \quad \text{and\quad \quad }\mathbf{r}%
^{\pm }=(r_{0}q^{\pm 1},r_{1}q^{\mp 2},r_{2})\;.
\end{equation}
The inclusion and projection map now are \cite{RW02} 
\begin{equation*}
\imath \,\left| n\right\rangle ^{+}=r_{0}\frac{q^{-2n+1}-r_{1}q^{-1}}{%
q-q^{-1}}\,\left| n\right\rangle \otimes \left| 0\right\rangle +\left|
n-1\right\rangle \otimes \left| 1\right\rangle \quad \text{and\quad }\tau
\,\left| n+1\right\rangle \otimes \left| 0\right\rangle =\left|
n\right\rangle ^{-}\;.
\end{equation*}
Again this representation theoretic fact is the platform for deriving a
functional equation of the type 
\begin{equation}
T(z)Q(z;\mathbf{r})=\psi _{1}(z)Q(z;\mathbf{r}^{+})+\psi _{2}(z)Q(z;\mathbf{r%
}^{-}),  \label{TQ2}
\end{equation}
where one now has 
\begin{equation}
Q(w;\mathbf{r})=(\limfunc{Tr}_{\mathbb{\pi }^{+}(w;\mathbf{r})}\otimes 1_{%
\mathcal{H}})\mathbf{Q}=\sum_{n=-\infty }^{\infty }Q_{nn}\;.  \label{Qgen}
\end{equation}
The reader will have noticed that the sum in the definition of the auxiliary
matrix is now infinite. In fact, this can cause convergence problems. From a
mathematical point of view one should therefore treat $q$ as an abstract
indeterminate (rather than a complex number) and view the matrix elements of
(\ref{Qgen}) as formal power series in $q$. We will return to this point
later when we calculate the spectrum of (\ref{Qgen}). See also the
discussion in \cite{RW02}.

\subsection{Commutation relations with the Yang-Baxter algebra}

Having specified explicitly the matrix elements of the intertwiner (\ref{L})
at and away from a root of unity we can exploit the Yang-Baxter equation (%
\ref{YBE}) in order to derive the commutation relations of the matrix
elements (\ref{Qmatrix}) with the generators (\ref{ABCD}) of the Yang-Baxter
algebra. One finds 
\begin{eqnarray}
\alpha _{k}Q_{kl}A &=&\alpha _{l}\,AQ_{kl}+\gamma _{l-1}\,BQ_{k\,l-1}-\beta
_{k}\,Q_{k-1\,l}C,  \label{QA} \\
\alpha _{k}Q_{kl}B &=&\delta _{l}\,BQ_{kl}+\beta _{l+1}\,AQ_{k\,l+1}-\beta
_{k}\,Q_{k-1\,l}\,D,  \label{QB} \\
\delta _{k}Q_{kl}C &=&\alpha _{l}\,CQ_{kl}+\gamma _{l-1}DQ_{kl-1}-\gamma
_{k}Q_{k+1\,l}A,  \label{QC} \\
\delta _{k}Q_{kl}D &=&\delta _{l}\,DQ_{kl}+\beta _{l+1}\,CQ_{k\,l+1}-\gamma
_{k}\,Q_{k+1\,l}B\;.  \label{QD}
\end{eqnarray}
Here we have suppressed the dependence on the spectral variables in the
notation which for say the second identity is explicitly given by 
\begin{multline*}
Q_{kl}(w)B(z)= \\
\frac{\delta _{l}(w/z)}{\alpha _{k}(w/z)}\,B(z)Q_{kl}(w)+\frac{\beta
_{l+1}(w/z)}{\alpha _{k}(w/z)}\,A(z)Q_{k\,l+1}(w)-\frac{\beta _{k\,}(w/z)}{%
\alpha _{k}(w/z)}\,Q_{k-1\,l}(w)\,D(z)\;.
\end{multline*}
Depending on the type of representation determining the auxiliary space of $%
Q $ there are different boundary conditions. For example, if $q^{N^{\prime
}}=\pm 1$ then 
\begin{eqnarray*}
\beta _{0} &=&0\;\Rightarrow \;\alpha _{0}Q_{0l}B=\delta _{l}\,BQ_{0l}+\beta
_{l+1}\,AQ_{0\,l+1}, \\
\beta _{N^{\prime }} &=&0\;\Rightarrow \;\alpha _{k}Q_{kN^{\prime
}-1}B=\delta _{N^{\prime }-1}\,BQ_{kN^{\prime }-1}-\beta
_{k}\,Q_{k-1\,N^{\prime }-1}\,D\;.
\end{eqnarray*}
Away from a root of unity the auxiliary space is infinite-dimensional and
there are no boundary conditions.

Notice at this point the difference in the alternative construction
procedures for auxiliary matrices. In order to obtain an auxiliary matrix
which commutes with the transfer matrix we used the concept of intertwiners
leading to (\ref{YBE}). The latter implies the commutator $[T(z),Q(w)]=0$
which corresponds to the following non-trivial equation in terms of the
Yang-Baxter algebra, 
\begin{equation}
\sum_{k}\left( \frac{\gamma _{k-1}}{\alpha _{k}}\,BQ_{k\,k-1}-\frac{\gamma
_{k}}{\delta _{k}}\,Q_{k+1\,k}B\right) =\sum_{k}\left( \frac{\beta _{k}}{%
\alpha _{k}}\,Q_{k-1\,k}C-\frac{\beta _{k+1}}{\delta _{k}}%
\,CQ_{k\,k+1}\right) \;.
\end{equation}
In contrast Baxter's method described e.g. in \cite{BxBook} relies on the
``pair propagation through a vertex'' property to construct an auxiliary
matrix which commutes with $T$. The relation (\ref{YBE}), which is the key
to relating the Yang-Baxter algebra (\ref{ABCD}) with the matrix elements (%
\ref{Qmatrix}), is missing in Baxter's approach.

\section{Action on Bethe states}

Since the transfer matrix and the $Q$-operators commute, they must allow for
a common set of eigenstates. This motivates us to compute the action of the $%
Q$-operator on the Bethe eigenstates of the transfer matrix, i.e. we now
make the crucial assumption that the quantum space $\mathcal{H}$ contains a
vector $\left| 0\right\rangle _{\mathcal{H}}$, called the pseudo-vacuum,
which satisfies 
\begin{equation}
A\left| 0\right\rangle _{\mathcal{H}}=\left| 0\right\rangle _{\mathcal{H}%
}\left\langle 0|A|0\right\rangle _{\mathcal{H}},\quad \quad C\left|
0\right\rangle _{\mathcal{H}}=0,\quad \quad D\left| 0\right\rangle _{%
\mathcal{H}}=\left| 0\right\rangle _{\mathcal{H}}\left\langle
0|D|0\right\rangle _{\mathcal{H}},
\end{equation}
and 
\begin{equation}
Q_{kk}\left| 0\right\rangle _{\mathcal{H}}=\left| 0\right\rangle _{\mathcal{H%
}}\left\langle 0|Q_{kk}|0\right\rangle _{\mathcal{H}},\quad \quad
Q_{jk}\left| 0\right\rangle _{\mathcal{H}}=0,\quad j>k\;.  \label{Qpseudo}
\end{equation}
In fact, when the quantum space $\mathcal{H}$ carries a representation $\pi
_{\mathcal{H}}$ of the quantum group $U_{q}(\widetilde{sl}_{2})$ or $%
U_{q}(b_{-})$ this pseudo-vacuum is identified with a (unique) highest
weight vector. The above properties then follow from the intertwining
property (\ref{inter}) of the monodromy matrices. For instance, one has for $%
\mathbf{Q}$ that 
\begin{equation*}
\lbrack \mathbf{Q},\pi _{w}^{\prime }(q^{h})\otimes \pi _{\mathcal{H}%
}(q^{h})]=0\quad \Rightarrow \quad \pi _{\mathcal{H}}(q^{h})Q_{jk}\pi _{%
\mathcal{H}}(q^{-h})=q^{2(j-k)}Q_{jk},
\end{equation*}
which then implies (\ref{Qpseudo}) by exploiting the fact that $\left|
0\right\rangle _{\mathcal{H}}$ is highest weight.

We now define a ``proper'' Bethe eigenstate as a vector of the form 
\begin{equation}
\left| z_{1},...,z_{n_{B}}\right\rangle _{\mathcal{H}}=%
\prod_{j=1}^{n_{B}}B(z_{j})\left| 0\right\rangle _{\mathcal{H}},
\label{Bethev}
\end{equation}
where the parameters $z_{j}=z_{j}(q,\lambda )$ are \emph{finite} solutions
to the ``generalized'' Bethe ansatz equations 
\begin{equation}
\left\langle 0|A(z_{i})|0\right\rangle _{\mathcal{H}%
}\,q^{n_{B}}P_{B}(z_{i}q^{-2})+\left\langle 0|D(z_{i})|0\right\rangle _{%
\mathcal{H}}\,q^{-n_{B}}P_{B}(z_{i}q^{2})=0\;.  \label{genBAE}
\end{equation}
Here we have introduced for convenience the ``Bethe polynomial'' 
\begin{equation}
P_{B}(z)=\prod_{j=1}^{n_{B}}(1-z/z_{j})\quad \text{with}\quad
z_{j}=z_{j}(q,\lambda )\;.  \label{PB}
\end{equation}
This set of coupled non-linear equations is sufficient to guarantee that the
Bethe states (\ref{Bethev}) are indeed eigenstates of the transfer matrix $%
T=A+D$. The proof follows the same lines as the well-known computation for
the XXZ spin-chain \cite{QISM}, 
\begin{equation}
T(z)=\left\langle 0|A(z)|0\right\rangle _{\mathcal{H}}\,q^{n_{B}}\frac{%
P_{B}(zq^{-2})}{P_{B}(z)}+\left\langle 0|D(z)|0\right\rangle _{\mathcal{H}%
}\,q^{-n_{B}}\frac{P_{B}(zq^{2})}{P_{B}(z)}\;.  \label{TABA}
\end{equation}

We restrict ourselves to \emph{finite} solutions of the Bethe ansatz
equations (\ref{genBAE}) in order to take into account certain peculiarities
which can occur at roots of unity. Some of the finite Bethe roots $z_{j}$ at 
$q^{N^{\prime }}\neq \pm 1$ can tend to zero or infinity when the root of
unity limit is taken. In this limit it might also occur that a subset of
Bethe roots $\{z_{i_{\ell }}\}_{\ell \in \mathbb{Z}_{N^{\prime }}}$ forms a
complete string, 
\begin{equation}
\lim_{q^{\prime }\rightarrow q}\prod_{\ell \in \mathbb{Z}_{N^{\prime
}}}(z-z_{i_{\ell }}(q^{\prime }))=\prod_{\ell \in \mathbb{Z}_{N^{\prime
}}}(z-z_{i_{0}}(q)q^{2\ell })=z^{N^{\prime }}-z_{i_{0}}(q)^{N^{\prime
}},\quad \quad q^{N^{\prime }}=\pm 1,
\end{equation}
and so drops out of the equation (\ref{genBAE}) \cite{FM01a}. In terms of
the Yang-Baxter algebra the occurence of a complete string corresponds to
the vanishing of the following operator product, 
\begin{equation}
\lim_{q^{\prime }\rightarrow q}\prod_{\ell \in \mathbb{Z}_{N^{\prime
}}}B(z_{i_{\ell }}(q^{\prime }))=0,\quad \quad q^{N^{\prime }}=\pm 1\;.
\end{equation}
Hence, the Bethe states (\ref{Bethev}) evaluated at a root of unity might
not yield a complete set of eigenstates. Moreover, the fact that the
transfer matrix and the $Q$-operator possess a common set of eigenvectors
does not necessarily imply that the Bethe states are also eigenstates of the 
$Q$-operator when degeneracies are present. This is precisely the case when $%
q^{N^{\prime }}=\pm 1$. For instance, in \cite{KQ}, auxiliary matrices have
been constructed for the six-vertex model at roots of unity which do not
preserve the total spin and whose eigenstates therefore are different from
the Bethe states. We will return to this discussion when we specialize our
general setup to the XXZ spin-chain. For the moment we keep the calculation
as general as possible.\smallskip

Let us start with the simplest case, only one ``Bethe root'' is present.
That is, we evaluate the $B$ operator at solutions $z_{0}$ to the equation 
\begin{equation}
\left\langle 0|A(z_{0})|0\right\rangle _{\mathcal{H}}\,=\left\langle
0|D(z_{0})|0\right\rangle _{\mathcal{H}}\;.
\end{equation}
In order to compute the action of the $Q$-operator on the corresponding
eigenstate of the transfer matrix we use the relation 
\begin{equation}
Q_{kl}B=\left( \tfrac{\delta _{l}}{\alpha _{k}}-\tfrac{\beta _{l+1}\gamma
_{l}}{\alpha _{k}\alpha _{l+1}}\right) BQ_{kl}+\tfrac{\beta _{l+1}}{\alpha
_{l+1}}\,Q_{kl+1}A-\tfrac{\beta _{k}}{\alpha _{k}}\,Q_{k-1\,l}D+\tfrac{\beta
_{l+1}\beta _{k}}{\alpha _{l+1}\alpha _{k}}\,Q_{k-1\,l+1}C  \label{QBBQ}
\end{equation}
which follows from the identities (\ref{QA}), (\ref{QB}). Note that if $k=l$
all of the rightmost operators on the right-hand-side of the above equation
possess the pseudo-vacuum as eigenvector. Taking the trace in (\ref{QBBQ})
we obtain (the boundary terms work out correctly if $q^{N^{\prime }}=\pm 1$) 
\begin{multline*}
\sum_{k}Q_{kk}B= \\
B\sum_{k}\left( \frac{\delta _{k}}{\alpha _{k}}-\frac{\beta _{k+1}\gamma _{k}%
}{\alpha _{k}\alpha _{k+1}}\right) Q_{kk}+\sum_{k}\frac{\beta _{k}}{\alpha
_{k}}\,Q_{k-1\,k}(A-D)+\sum_{k}\frac{\beta _{k+1}\beta _{k}}{\alpha
_{k}\alpha _{k+1}}\,Q_{k-1\,k+1}\,C\;.
\end{multline*}
When acting on the pseudovacuum the second and third term on the RHS of the
equation vanish, which leaves us with the eigenvalue 
\begin{equation}
Q(w)B(z_{0})\left| 0\right\rangle _{\mathcal{H}}=\left\{ \sum_{k}\left( 
\frac{\delta _{k}}{\alpha _{k}}-\frac{\beta _{k+1}\gamma _{k}}{\alpha
_{k}\alpha _{k+1}}\right) \,\left\langle 0|Q_{kk}|0\right\rangle _{\mathcal{H%
}}\right\} \,B(z_{0})\left| 0\right\rangle _{\mathcal{H}}\;.
\end{equation}
Here the sum runs over $\mathbb{Z}_{N^{\prime }}$ when $q^{N^{\prime }}=\pm
1 $ and over $\mathbb{Z}$ when $q$ is not a root of unity. The calculation
for Bethe states with multiple Bethe roots follows the same logic, though it
is now much more involved to show that all of the ``unwanted'' terms vanish
due to the Bethe ansatz equations; see the appendix.

\subsection{Conjecture}

For a Bethe state with $n_{B}$ Bethe roots we conjecture the following
formula for the eigenvalues of the respective auxiliary matrices: 
\begin{eqnarray}
Q(w)\prod_{j=1}^{n_{B}}B(z_{j})\left| 0\right\rangle _{\mathcal{H}} &=&
\label{Con} \\
&&\hspace{-3.5cm}\left\{ \sum_{k}\,\left\langle 0|Q_{kk}(w)|0\right\rangle _{%
\mathcal{H}}\prod_{j=1}^{n_{B}}\left( \frac{\delta _{k}(w/z_{j})}{\alpha
_{k}(w/z_{j})}-\frac{\beta _{k+1}(w/z_{j})\gamma _{k}(w/z_{j})}{\alpha
_{k}(w/z_{j})\alpha _{k+1}(w/z_{j})}\right) \right\}
\,\prod_{j=1}^{n_{B}}B(z_{j})\left| 0\right\rangle _{\mathcal{H}}  \notag
\end{eqnarray}
Note that the particular choice of the quantum space only enters via the
pseudo-vacuum expectation value $\left\langle 0|Q_{kk}|0\right\rangle _{%
\mathcal{H}}$ and the Bethe roots $\{z_{j}\}_{j=1}^{n_{B}}.$ The combination
of matrix elements appearing in the product of the eigenvalue expression
only depends on the auxiliary space. The above conjecture can be verified
for states with $n_{B}=2,3$ by employing the intermediate steps detailed in
the appendix. For $n_{B}>3$ we will provide further support by showing that
the eigenvalues satisfy the functional equations (\ref{TQ1}) and (\ref{TQ2})
respectively. For this purpose we need to determine the coefficient
functions in (\ref{TQ1}) and (\ref{TQ2}) first. The latter depend on the
choice of the quantum space and we now specialize to the XXZ spin-chain.

\section{The XXZ spin-chain}

As remarked upon earlier, one chooses for the familiar case of the
inhomogeneous six-vertex model or XXZ spin-chain with quasi-periodic
boundary conditions, 
\begin{equation}
\pi _{\mathcal{H}}=\bigotimes_{m=1}^{M}\pi _{\zeta _{m}}^{(1)}\quad \quad 
\text{and\quad \quad }\mathcal{H}=(\mathbb{C}^{2})^{\otimes M}\;.
\label{XXZchain}
\end{equation}
Here $\{\zeta _{m}\}$ are some complex inhomogeneity parameters. The
transfer and auxiliary matrix can then be written as the trace of an
operator product 
\begin{equation}
T(z)=\limfunc{Tr}_{\pi _{z}^{(1)}}\lambda ^{\sigma ^{z}\otimes
1}R_{0M}(z/\zeta _{M})\cdots R_{01}(z/\zeta _{1}),
\end{equation}
and 
\begin{equation}
Q(w)=\limfunc{Tr}_{\pi _{w}^{\prime }}\lambda ^{\pi _{w}^{\prime }(h)\otimes
1}L_{0M}(w/\zeta _{M})\cdots L_{01}(w/\zeta _{1})\;.  \label{T&Q6v}
\end{equation}
Here $R$ and $L$ are the $R$-matrix and the $L$-operator specified earlier.
Notice again the difference with Baxter's construction procedure \cite
{BxBook}. His final $Q,$ which commutes with the transfer matrix, is in
general not of this simple form. For the pseudo-vacuum and the associated
``expectation values'' one finds 
\begin{equation}
\left| 0\right\rangle _{\mathcal{H}}=\left| 0\right\rangle _{\mathbb{C}%
^{2}}\otimes \cdots \otimes \left| 0\right\rangle _{\mathbb{C}^{2}},\quad
\left\langle 0|A|0\right\rangle _{\mathcal{H}}=\lambda ,\quad \left\langle
0|D|0\right\rangle _{\mathcal{H}}=\lambda ^{-1}q^{M}\prod_{m=1}^{M}\frac{%
z-\zeta _{m}}{zq^{2}-\zeta _{m}}\;,
\end{equation}
and 
\begin{equation}
\left\langle 0|\mathbf{Q}(w)|0\right\rangle _{\mathcal{H}}=\lambda ^{\pi
_{w}^{\prime }(h^{\prime })\otimes 1}\prod_{m=1}^{M}\mathbf{\alpha }(w/\zeta
_{m})\;\Rightarrow \;Q_{nn}\left| 0\right\rangle _{\mathcal{H}}=\lambda
^{-2n}\prod_{m=1}^{M}\alpha _{n}(w/\zeta _{m})\left| 0\right\rangle _{%
\mathcal{H}}\;.
\end{equation}
Here we have slightly modified our earlier conventions. Instead of taking
the Cartan generator $h$ in the exponent of the twist parameter, we
introduced for convenience $h^{\prime }$ which in the two representations (%
\ref{pi1}) and (\ref{pi2}) is given by 
\begin{equation}
\pi _{w}^{\mu }(h^{\prime })\left| n\right\rangle =-2n\left| n\right\rangle
\quad \text{and\quad }\pi ^{+}(h^{\prime })\left| n\right\rangle =-2n\left|
n\right\rangle \;.
\end{equation}
This simply amounts to a renormalization of the respective auxiliary
matrices by an overall factor. We continue to treat the two cases of $q$
being a root of unity and $q$ not being a root of unity separately.

\subsection{Roots of unity}

When $q^{N^{\prime }}=\pm 1$, the auxiliary space is finite and upon
inserting the expressions (\ref{G1}) into the conjectures formula (\ref{Con}%
) one obtains 
\begin{equation}
\left\langle 0|Q_{nn}(w)|0\right\rangle _{\mathcal{H}}=\lambda ^{-2n}\mu ^{%
\frac{M}{2}}q^{Mn+\frac{M}{2}}\prod_{m=1}^{M}(w\mu ^{-1}q^{-2n}/\zeta
_{m}-1),
\end{equation}
and 
\begin{eqnarray}
Q_{\mu }(w)\prod_{j=1}^{n_{B}}B(z_{j})\left| 0\right\rangle _{\mathcal{H}}
&=&  \label{Qbethe} \\
&&\hspace{-4.5cm}\left\{ q^{S^{z}}\mu ^{S^{z}}P_{B}(w\mu )P_{B}(w\mu
^{-1})\sum_{k\in \mathbb{Z}_{N^{\prime }}}\frac{\lambda
^{-2k}q^{2kS^{z}}\prod_{m=1}^{M}(w\mu ^{-1}q^{-2k}/\zeta _{m}-1)}{P_{B}(w\mu
^{-1}q^{-2k})P_{B}(w\mu ^{-1}q^{-2k-2})}\right\}
\prod_{j=1}^{n_{B}}B(z_{j})\left| 0\right\rangle _{\mathcal{H}}\;.  \notag
\end{eqnarray}
Here we have used the relation between total spin and the number of Bethe
roots, $2S^{z}=M-2n_{B}$. We emphasize that the Bethe roots $z_{j}$ are
assumed to be \emph{finite} solutions to the Bethe ansatz equations \emph{at 
}a root of unity. This is different from the polynomial which is obtained by
solving the Bethe ansatz equations away from a root of unity and then taking
the root of unity limit. As pointed out before, in this limit one might
encounter vanishing or infinite Bethe roots as well as complete strings.
Before we investigate these issues we first check the eigenvalues of the
auxiliary matrix for consistency by making contact with the functional
equation (\ref{TQ1}).

\subsubsection{The $TQ$-equation}

Using the explicit form of the inclusion and projection map one calculates
for the twisted XXZ spin-chain the following coefficient functions in (\ref
{TQ1}), 
\begin{equation}
T(z)Q_{\mu }(z/\mu )=\lambda ^{-1}q^{\frac{M}{2}}\left( \prod_{m=1}^{M}\frac{%
z-\zeta _{m}}{zq^{2}-\zeta _{m}}\right) Q_{\mu q}(zq/\mu )+\lambda q^{\frac{M%
}{2}}Q_{\mu q^{-1}}(zq^{-1}/\mu )\;.  \label{TQ6v1}
\end{equation}
Employing the known expression (\ref{TABA}) for the eigenvalues of the
transfer matrix within the framework of the algebraic Bethe ansatz \cite
{QISM}, 
\begin{equation}
T(z)=\lambda q^{n_{B}}\frac{P_{B}(zq^{-2})}{P_{B}(z)}+\lambda ^{-1}\left(
\prod_{m=1}^{M}\frac{z-\zeta _{m}}{zq^{2}-\zeta _{m}}\right) q^{M-n_{B}}%
\frac{P_{B}(zq^{2})}{P_{B}(z)}\;,  \label{Taba6v}
\end{equation}
one verifies that the conjectured eigenvalues of the auxiliary matrix
satisfy (\ref{TQ6v1}). Notice that we have implictly made the assumption
that all $Q$-matrices in (\ref{TQ6v1}) commute with each other. For $\lambda
=1$ this has been proven in \cite{KQ}. Using the explicit form of the
intertwiners employed in this proof one verifies that the same holds true
for quasi-periodic boundary conditions as long as $\lambda ^{N}=1$. This is
sufficient to cover the symmetries investigated in \cite{Ktw}.

\subsubsection{The degeneracies at roots of unity $q^{N^{\prime }}=\pm 1$}

Recall from \cite{DFM,Ktw} that due to the loop symmetry at a root of unity $%
q^{N^{\prime }}=\pm 1$ the eigenspaces of the transfer matrix are organized
into multiplets containing states whose total spin $S^{z}$ varies by
multiples of $N^{\prime }$. The Bethe states (\ref{Bethev}) with finite
Bethe roots are assumed to correspond to the highest weight states in such
multiplets, i.e. acting with the generators of the respective symmetry
algebra on this state one successively obtains the whole degenerate
eigenspace of the transfer matrix. (See \cite{FM01b,D02,KQ2}\ for examples
of the highest weight property.) The states within a degenerate multiplet
are \emph{not} of the form (\ref{Bethev}), whence our result (\ref{Qbethe})
only applies to the highest weight vectors containing finite Bethe roots.
Therefore, it does not yield the complete spectrum of the auxiliary matrices 
$Q_{\mu }$. We now compare for these highest weight states the identity (\ref
{Qbethe}) with the formulae (11) and (19) in \cite{KQ2} which have been
proven for $N=3$ using functional relations and conjectured to hold true for 
$N>3$ based on numerical data.

Recall from \cite{KQ2} that the eigenvalues of the auxiliary matrices $%
Q_{\mu }$ for the homogeneous chain, $\{\zeta _{m}=1\}_{m=1}^{M},$ and
periodic boundary conditions, $\lambda =1$, were shown to be of the
following general form (see formulae (11) and (19) in \cite{KQ2}), 
\begin{equation}
Q_{\mu }(w=z/\mu )=\mathcal{N}_{\mu }\,z^{\bar{n}_{\infty
}}P_{B}(z)P_{B}(z\mu ^{-2})P_{S}(z^{N^{\prime }},\mu ^{2N^{\prime }})\;.
\label{19}
\end{equation}
In order to match our result (\ref{Qbethe}) from the algebraic Bethe ansatz
computation with the formulae (11) and (19) in \cite{KQ2} we have to
identify 
\begin{equation}
\mathcal{N}_{\mu =1}P_{S}(z^{N^{\prime }},\mu =1)=\mathcal{N}_{\mu
=1}\prod_{j=1}^{n_{S}}(1-z^{N^{\prime }}/a_{j})=q^{S^{z}}\sum_{k\in \mathbb{Z%
}_{N^{\prime }}}\frac{q^{2kS^{z}}(zq^{-2k}-1)^{M}}{%
P_{B}(zq^{-2k})P_{B}(zq^{-2k-2})}\;.  \label{PS}
\end{equation}
Note that in our derivation of (\ref{Qbethe}) we have assumed infinite and
vanishing Bethe roots to be absent. Thus, we have to set $\bar{n}_{\infty
}=0 $ in formula (19) of \cite{KQ2}. Numerical evidence suggests that this
is only true in the commensurate sectors $2S^{z}=0\func{mod}N$. One then
finds for $N=3$ agreement between the results in \cite{KQ2} and our present
calculation up to a trivial redefinition of the normalization constant $%
\mathcal{N}_{\mu }$. Recall from the discussion in \cite{KQ2} that the
rational function (\ref{PS}) is in fact a polynomial\footnote{%
We briefly recall the argument. First, notice that the Bethe ansatz
equations (\ref{genBAE}) ensure that the rational function (\ref{PS}) is
pole free in $z$. Hence, it must be a polynomial in $z$. Secondly, the
function (\ref{PS}) is obviously invariant under the replacement $%
z\rightarrow zq^{2}$. Thus, any zero $z^{\prime }\neq 0$ occurs inside a
string $\{z^{\prime }q^{2\ell }\}_{\ell \in \mathbb{Z}_{N^{\prime }}}$ and
gives rise to a factor $(z^{N^{\prime }}-z^{\prime N^{\prime }})$ in the
root decomposition of the polynomial.} in the variable $z^{N^{\prime }}$
whose roots $a_{j}$ contain all the essential information on the irreducible
representation of the loop algebra spanning the degenerate eigenspaces of
the transfer matrix. The reader is referred to \cite{KQ2} for details.

In contrast to the transfer matrix the auxiliary matrices $Q_{\mu }$ stay
non-degenerate and in addition to the eigenvalues corresponding to the
highest weight state one also needs to compute the remaining eigenvalues of
the auxiliary matrix within the multiplet. The examples explicitly worked
out for $N=3$ and $M=5,6,8$ in \cite{KQ2} suggest the following picture:

Set $\lambda =1,\;\{\zeta _{m}=1\}_{m=1}^{M}$ and assume we have a multiplet
in a commensurate sector, i.e. the highest weight state has spin $2S^{z}=0%
\func{mod}N$. Then the form of the eigenvalue (\ref{Qbethe}) remains the
same except for a change in the $\mu $-dependence of the polynomial (\ref{PS}%
). The examples in \cite{KQ2} showed that this polynomial changes within the
multiplet according to the following rule, 
\begin{eqnarray*}
S^{z} &:&\quad P_{S}(z^{N^{\prime }},\mu ^{2N^{\prime
}})=\prod_{j=1}^{n_{S}}(1-z^{N^{\prime }}\mu ^{-2N^{\prime }}/a_{j})\quad
\quad \text{(highest weight state)} \\
S^{z}-N^{\prime } &:&\quad P_{S}(z^{N^{\prime }},\mu ^{2N^{\prime
}})=(1-z^{N^{\prime }}/a_{1})\prod_{j=2}^{n_{S}}(1-z^{N^{\prime }}\mu
^{-2N^{\prime }}/a_{j}) \\
S^{z}-2N^{\prime } &:&\quad P_{S}(z^{N^{\prime }},\mu ^{2N^{\prime
}})=(1-z^{N^{\prime }}/a_{1})(1-z^{N^{\prime
}}/a_{2})\prod_{j=3}^{n_{S}}(1-z^{N^{\prime }}\mu ^{-2N^{\prime }}/a_{j}) \\
&&\quad \quad \quad \quad \vdots \\
-S^{z} &:&\quad P_{S}(z^{N^{\prime }})=\prod_{j=1}^{n_{S}}(1-z^{N^{\prime
}}/a_{j})\quad \quad \text{(lowest weight state)}
\end{eqnarray*}
That is, as one steps through the multiplet the polynomial (\ref{PS})
successively looses factors of $\mu ^{-2N^{\prime }}$ until one reaches the
lowest weight state, where it does not contain any $\mu $-factor. We can
confirm this picture by computing the eigenvalue corresponding to the lowest
weight state. Invoking the transformation of the auxiliary matrix under spin
reversal $\frak{R}=\prod_{m=1}^{M}\sigma _{m}^{x}$ one finds (cf formula
(39) in \cite{KQ2}), 
\begin{multline}
Q_{\mu }(w=z/\mu )\prod_{j=1}^{n_{B}}C(z_{j})\frak{R}\left| 0\right\rangle _{%
\mathcal{H}}=\text{const.}\times \frak{R}Q_{\mu ^{-1}}(w=z/\mu
)\prod_{j=1}^{n_{B}}B(z_{j})\left| 0\right\rangle _{\mathcal{H}}=  \notag \\
\left\{ q^{-S^{z}}\mu ^{S^{z}}P_{B}(z)P_{B}(z\mu ^{-2})\sum_{k\in \mathbb{Z}%
_{N^{\prime }}}\lambda ^{-2k}q^{2kS^{z}}\frac{\prod_{m=1}^{M}(zq^{-2k}/\zeta
_{m}-1)}{P_{B}(zq^{-2k})P_{B}(zq^{-2k-2})}\right\}
\prod_{j=1}^{n_{B}}C(z_{j})\frak{R}\left| 0\right\rangle _{\mathcal{H}}\;.
\end{multline}
In accordance with the picture outlined before, the sum yielding the
polynomial $P_{S}$ is now independent of the parameter $\mu $.

\subsubsection{Infinite and vanishing Bethe roots}

Besides the missing states within a degenerate multiplet of the transfer
matrix, we are also missing those states which contain infinite or vanishing
Bethe roots in the root of unity limit. A concrete example for the
homogeneous chain, i.e. $\{\zeta _{m}=1\}_{m=1}^{M},$ is given by setting $%
M=5,\;S^{z}=1/2$ and considering the subspace of momentum $P=0$. There one
finds for $2\Delta =q^{\prime }+q^{\prime -1}$ with $q^{\prime }$ not a root
of unity the following Bethe roots 
\begin{equation*}
b_{1}=b_{2}^{-1}=\frac{1}{4}\left( 1+\Delta -\sqrt{5+\Delta (\Delta -2)}+%
\sqrt{\left( 1+\Delta -\sqrt{5+\Delta (\Delta -2)}\right) ^{2}-16}\right) \;.
\end{equation*}
Here we have temporarily introduced the parameterization $b_{i}=q^{\prime
}(1-z_{i})/(1-z_{i}q^{\prime 2})$ in order to accommodate the occurrence of
infinite roots. Taking the root of unity limit $q^{\prime }\rightarrow
q=e^{2\pi i/3}$ one now computes 
\begin{equation*}
\lim_{q^{\prime }\rightarrow q}T(z)=1+\left( \frac{q(1-z)}{1-zq^{2}}\right)
^{5},\quad \lim_{q^{\prime }\rightarrow q}b_{1}=\lim_{q^{\prime }\rightarrow
q}b_{2}^{-1}=e^{2\pi i/3}=q,\quad z_{1}=0,\quad z_{2}=\infty \;.
\end{equation*}

Therefore the \emph{finite} solutions to the Bethe ansatz equations do not
give a complete set of eigenstates at roots of unity. Notice that it might
even happen that the highest weight state in a multiplet contains infinite
or vanishing Bethe roots.

\subsection{The case when $q$ is not a root of unity}

A priori one might expect that things are now simpler than the root of unity
case as the only degeneracies which occur in the spectrum of the transfer
matrix are due to spin-reversal symmetry. However, the auxiliary space is
now infinite and one has to deal with potential convergence problems when
taking the trace in (\ref{Qgen}). This becomes evident if we insert the
expressions (\ref{G2}) into the conjectured formula for the eigenvalue
expression of the auxiliary matrix. One obtains for the pseudo-vacuum
expectation value 
\begin{equation}
\left\langle 0|Q_{nn}(w)|0\right\rangle _{\mathcal{H}}=\lambda
^{-2n}\prod_{m=1}^{M}\alpha _{n}(w/\zeta _{m})=\lambda ^{-2n}r_{0}^{-\frac{M%
}{2}}q^{n\frac{M}{2}}\prod_{m=1}^{M}(wr_{2}q^{-2n+2}/\zeta _{m}-1)
\end{equation}
and the action of the auxiliary matrix on a Bethe state produces 
\begin{multline}
Q(z;r_{0},r_{1},r_{2})\prod_{j=1}^{n_{B}}B(z_{j})\left| 0\right\rangle _{%
\mathcal{H}}=  \notag \\
\left\{ r_{0}^{-S^{z}}P_{B}(z)P_{B}(zr_{1}r_{2})\sum_{\ell \in \mathbb{Z}}%
\frac{\lambda ^{-2\ell }q^{2\ell S^{z}}\prod_{m=1}^{M}(zr_{2}q^{-2\ell
+2}/\zeta _{m}-1)}{P_{B}(zr_{2}q^{-2\ell +2})P_{B}(zr_{2}q^{-2\ell })}%
\right\} \prod_{j=1}^{n_{B}}B(z_{j})\left| 0\right\rangle _{\mathcal{H}}\;.
\end{multline}
The above expression for the eigenvalue might not be convergent. In fact, it
was discussed in \cite{RW02} for $r_{1}=r_{2}=0$ and $\lambda =1$ that the
matrix elements of the $Q$-operator contain the formal power series 
\begin{equation}
\delta (q^{2S^{z}})=\sum_{\ell \in \mathbb{Z}}q^{2\ell S^{z}},\quad \quad
S^{z}=\frac{M}{2}-n_{B}  \label{delta}
\end{equation}
in the deformation parameter $q$. If one removes this factor in the sector $%
S^{z}=0$ by an ad-hoc renormalisation the remaining auxiliary matrix can be
identified with Baxter's expression (101) in \cite{Bx73a}. See the
discussion in Section 5 of \cite{RW02} for details. Obviously, our algebraic
Bethe ansatz analysis of the auxiliary matrices reproduces these findings.
In the limit $r_{1},r_{2}\rightarrow 0$ we obtain apart from the factor (\ref
{delta}) and the constant $r_{0}^{-S^{z}},$ the Bethe polynomial (\ref{PB})
as eigenvalue as we should. However, the required renormalisation is
certainly an unsatisfactory answer to the convergence problem, in particular
as the factor (\ref{delta}) is needed outside the spin-sector $S^{z}=0$ in
order to satisfy the functional equation (\ref{TQ2}) which for the XXZ
spin-chain is calculated to 
\begin{eqnarray}
Q(z;r_{0},r_{1},r_{2})T(z) &=&  \label{TQ6v2} \\
&&\hspace{-3.5cm}\lambda ^{-1}q^{\frac{M}{2}%
}Q(zq^{-2};r_{0}q^{-1},r_{1}q^{2},r_{2})+\lambda q^{\frac{M}{2}}\left(
\prod_{m=1}^{M}\frac{z-\zeta _{m}}{zq^{2}-\zeta _{m}}\right)
Q(zq^{2};r_{0}q,r_{1}q^{-2},r_{2})\;.  \notag
\end{eqnarray}
In order to obtain a convergent expression let us modify the auxiliary space
by restricting it to an invariant subspace at special values of the
parameters $r_{1,2}$.

\subsubsection{Truncation of the auxiliary space}

In order to establish convergence it would be helpful to restrict the sum in
the eigenvalue expression of the auxiliary matrix either to the positive or
negative integers and then choose the twist parameter appropriately. In
order to achieve this we have to find values for the free parameters $%
(r_{0},r_{1},r_{2})$ such that the representation (\ref{pi2}) contains an
invariant subspace. To this end we need for some $m_{o}\in \mathbb{Z}$ the
truncation condition 
\begin{equation}
e_{0}\left| m_{o}\right\rangle _{\pi ^{+}}=w\;\frac{%
r_{1}r_{2}+1-r_{1}q^{2m_{o}}-r_{2}q^{-2m_{o}}}{(q-q^{-1})^{2}}\left|
m_{o}\right\rangle _{\pi ^{+}}=0\;.  \label{truncc}
\end{equation}
This is easily achieved by setting 
\begin{equation}
\text{either\quad }r_{1}=q^{-2m_{o}}\text{\quad or\quad }r_{2}=q^{2m_{o}}\;.
\end{equation}
For simplicity let us choose $m_{o}=0$ and $r_{2}=1$. Denote by $\pi _{\leq
}^{+}=\pi _{\leq }^{+}(w;r_{0},r_{1})$ the irreducible subrepresentation of (%
\ref{pi2}) which is obtained by restriction to the negative integers
(including zero) and set 
\begin{equation}
Q_{\leq }(w;r_{0},r_{1})=(\limfunc{Tr}_{\mathbb{\pi }_{\leq }^{+}}\otimes 1_{%
\mathcal{H}})\mathbf{Q}=\sum_{n=-\infty }^{0}Q_{nn}\;.
\end{equation}
Since the parameter $r_{2}$ does not shift in the decomposition (\ref{seq2})
this truncation is consistent with the functional equation (\ref{TQ2}),
albeit we have to supplement the projection map $\tau $ by defining 
\begin{equation}
\tau \,\left| 0\right\rangle _{\pi _{\leq }^{+}(w;r_{0},r_{1})}\otimes
\left| 1\right\rangle _{\pi _{z}^{(1)}}\equiv \frac{r_{0}(r_{1}-1)}{q^{2}-1}%
\,\left| 0\right\rangle _{\pi _{\leq
}^{+}(wq^{-2};r_{0}q^{-1},r_{1}q^{2})}\quad r_{1}\neq 1,\quad w=z\;.
\end{equation}
For the semi-infinite representation space we now obtain the following
eigenvalue associated with a Bethe state (\ref{Bethev}), 
\begin{equation}
Q_{\leq }(z;r_{0},r_{1})=\lambda
^{-2}q^{2S^{z}}r_{0}^{-S^{z}}P_{B}(zr_{1})P_{B}(z)\sum_{\ell =1}^{\infty
}\lambda ^{2\ell }q^{-2\ell S^{z}}\frac{\prod_{m=1}^{M}(zq^{2\ell }/\zeta
_{m}-1)}{P_{B}(zq^{2\ell })P_{B}(zq^{2\ell -2})},  \label{Qtrunc}
\end{equation}
Here we have denoted the eigenvalue by the same symbol as the operator.
Unlike in the previous expression the sum now only ranges over the positive
integers. This puts us in the position to prove absolute convergence of the
series in (\ref{Qtrunc}). We treat the cases $|q|=1$ and $|q|^{\pm 1}>1$
separately.

Suppose $q$ is of modulus one, then we deduce from 
\begin{equation}
|q|=1:\quad \sum_{\ell =1}^{\infty }\left| \lambda ^{2\ell }q^{-2\ell S^{z}}%
\frac{\prod_{m=1}^{M}(zq^{2\ell }/\zeta _{m}-1)}{P_{B}(zq^{2\ell
})P_{B}(zq^{2\ell -2})}\right| \leq \frac{\prod_{m=1}^{M}(|z/\zeta _{m}|+1)}{%
\prod_{j=1}^{n_{B}}(1-|z/z_{j}|)^{2}}\sum_{\ell =1}^{\infty }|\lambda
|^{2\ell }
\end{equation}
that absolute convergence is guaranteed as long as $|\lambda |<1$.

Now let $|q|^{\pm 1}>1$ and set $q=\exp v,\;z=\exp 2u,\;\zeta _{m}=\exp 2\xi
_{m}$. Rewriting 
\begin{multline*}
\sum_{\ell =1}^{\infty }\left| \lambda ^{2\ell }q^{-2\ell S^{z}}\frac{%
\prod_{m=1}^{M}(zq^{2\ell }/\zeta _{m}-1)}{P_{B}(zq^{2\ell })P_{B}(zq^{2\ell
-2})}\right| = \\
\sum_{\ell =1}^{\infty }\frac{2|\lambda |^{2\ell }|q|^{n_{B}}|z|^{\frac{M}{2}%
-n_{B}}\prod_{m=1}^{M}|\sinh (u-\xi _{m}+\ell v)||\zeta _{m}|^{-\frac{1}{2}}%
}{\prod_{j=1}^{n_{B}}|\sinh (u-u_{j}+\ell v)\sinh (u-u_{j}+(\ell
-1)v)|/|z_{j}|}\leq \text{const.}\sum_{\ell =1}^{\infty }|\lambda |^{2\ell
}|q|^{\pm 2\ell |S^{z}|}
\end{multline*}
we find absolute convergence for $|\lambda |<|q|^{\mp M/2}$. Thus, in
summary we are left with the condition 
\begin{equation}
|\lambda |<|q|^{\mp M/2},\text{\quad }|q|^{\pm 1}\geq 1  \label{qconv}
\end{equation}
which now includes the case when $q$ is of modulus one.

Having assured convergence we can now verify that the eigenvalues (\ref
{Qtrunc}) satisfy the functional equation (\ref{TQ6v2}). Doing so we again
implicitly made the assumption that the $Q$-operators in the functional
equation commute with each other. Since the auxiliary space is still
semi-infinite, the proof of commutation via the construction of the
corresponding intertwiners (similar to the root of unity case) is less
feasible. Instead we are going to exploit the completeness of the Bethe
ansatz when $q$ is not a root of unity.

Provided one accepts that the Bethe states (\ref{Bethev}) and their
counterparts under spin-reversal 
\begin{equation}
\frak{R}\left| z_{1},...,z_{n_{B}}\right\rangle _{\mathcal{H}}\propto
\prod_{j=1}^{n_{B}}C(z_{j})\frak{R}\left| 0\right\rangle _{\mathcal{H}%
}\,,\quad \frak{R}\left| 0\right\rangle _{\mathcal{H}}=\left| 1\right\rangle
_{\mathbb{C}^{2}}\otimes \cdots \otimes \left| 1\right\rangle _{\mathbb{C}%
^{2}}\,,  \label{Rbethev}
\end{equation}
form a complete set of eigenstates which span the whole quantum space $%
\mathcal{H}$, one deduces that the three-parameter family $\{Q_{\leq
}(z;r_{0},r_{1})\}$ of auxiliary matrices can be simultaneously diagonalized
as the Bethe roots only depend on $q$ and $\lambda $. Hence, we must have $%
[Q_{\leq }(z;r_{0},r_{1}),Q_{\leq }(w;r_{0}^{\prime },r_{1}^{\prime })]=0$.

Notice that the commutation of the $Q$-operators for arbitrary spectral
parameters together with the explicit form of the matrix elements (\ref{G2})
implies that the eigenvalues (\ref{Qtrunc}) are polynomials of degree $\leq
M $ in the spectral variable $z$. Unlike the root-of-unity case, however,
the sum by itself is not a polynomial. One now has to include a factor $%
P_{B} $ in front of the sum in order to cancel the simple poles of the
denominator at $z=z_{j}$. Employing the Bethe ansatz equations (\ref{genBAE}%
) one then deduces that for any closed contour $C_{\ell }$ around $%
z_{j}q^{-2\ell }$ with integer $\ell \geq 0$ one has 
\begin{equation*}
\oint_{C_{\ell }}dz\;P_{B}(z)\sum_{\ell =1}^{\infty }\lambda ^{2\ell
}q^{-2\ell S^{z}}\frac{\prod_{m=1}^{M}(zq^{2\ell }/\zeta _{m}-1)}{%
P_{B}(zq^{2\ell })P_{B}(zq^{2\ell -2})}=0\;.
\end{equation*}
The vanishing of the contour integral signals that the above rational
function is indeed a polynomial in $z$. This is consistent with (\ref{Qtrunc}%
), where the parameter $r_{1}$ in the argument of the first factor is
arbitrary.

For completeness we briefly discuss the transformation under spin-reversal
in order to obtain the eigenvalues of the auxiliary matrix ``beyond the
equator'', i.e. for the states (\ref{Rbethev}). Acting with the
spin-reversal operator on the auxiliary matrix we obtain 
\begin{equation*}
\frak{R}Q_{\leq }(z;r_{0},r_{1})\frak{R}=\limfunc{Tr}_{\pi _{\leq
}^{+}}\lambda ^{\pi _{\leq }^{+}(h^{\prime })\otimes 1}\sigma
_{M}^{x}L_{0M}(z/\zeta _{M})\sigma _{M}^{x}\cdots \sigma
_{1}^{x}L_{01}(z/\zeta _{1})\sigma _{1}^{x}\;.
\end{equation*}
which amounts in terms of the intertwiner $L$ to the replacement 
\begin{equation}
\{\mathbf{\alpha },\mathbf{\beta },\mathbf{\gamma },\mathbf{\delta }%
\}\rightarrow \{\mathbf{\delta },\mathbf{\gamma },\mathbf{\beta },\mathbf{%
\alpha }\}\;.
\end{equation}
From (\ref{G2}) one derives the following identities for the matrix elements
of the intertwiner $L$, 
\begin{eqnarray}
\alpha _{n}(wq^{-2},q;r_{0},r_{1},r_{2}) &=&\delta
_{n}(w,q^{-1};r_{0}^{-1},r_{2},r_{1})  \notag \\
\delta _{n}(wq^{-2},q;r_{0},r_{1},r_{2}) &=&\alpha
_{n}(w,q^{-1};r_{0}^{-1},r_{2},r_{1})  \notag \\
\beta _{n}\gamma _{n-1}|_{(wq^{-2},q;r_{0},r_{1},r_{2})} &=&\beta _{n}\gamma
_{n-1}|_{(w,q^{-1};r_{0}^{-1},r_{2},r_{1})}
\end{eqnarray}
Notice that only the above combinations of matrix elements contribute to the
trace of the monodromy matrix (\ref{momQ}), i.e. the auxiliary matrix (\ref
{T&Q6v}). Hence, we have the identity 
\begin{equation}
\frak{R}Q_{\leq }(zq^{-2},q;r_{0},r_{1},r_{2}=1)\frak{R}=Q_{\leq
}(z,q^{-1};r_{0}^{-1},r_{2}=1,r_{1})\;.  \label{RQ}
\end{equation}
Thus, spin-reversal leads to the consideration of the truncated auxiliary
matrix with $r_{1}=1$ and $r_{2}$ arbitrary. We now discuss this case in the
context of $q$-oscillator representations. Before we have the
following\smallskip

\textbf{Remark}. In order to ensure convergence for $|q|=1$ we have excluded
the quasi-periodic boundary conditions with $|\lambda |\geq 1$. However, in
order to recover for example periodic boundary conditions, $\lambda =1,$ one
may proceed as follows. Employing (\ref{TQ6v2}) one can express the transfer
matrix eigenvalues in terms of the auxiliary matrix at $|\lambda |<1$. While
at the moment we do not have an analytic argument due to the implicit
dependence of the Bethe roots on $\lambda ,$ numerical calculations suggest
that one can then safely take the limit $\lambda \rightarrow 1$ in order to
recover the eigenvalues of the six-vertex transfer matrix with periodic
boundary conditions.

\subsubsection{$q$-oscillator representations}

We now simplify the representation (\ref{pi2}) of \cite{RW02} further in
order to make contact with the results in \cite{BLZ99} for the Coulomb gas
formalism of conformal field theory. We explicitly show that one can derive
the analogue of the functional equations therein also for the XXZ spin-chain.

The $q$-oscillator algebra is defined in terms of the generators \cite{Kul} 
\begin{equation}
qe_{+}e_{-}-q^{-1}e_{-}e_{+}=(q-q^{-1})^{-1}\quad \text{and\quad }[h,e_{\pm }%
]=\pm 2e_{\pm }\;.  \label{qos}
\end{equation}
We define two representations $\varrho _{\pm }$ by identifying $%
e_{0}\rightarrow e_{\pm },\;e_{1}\rightarrow e_{\mp
},\;h_{1}=-h_{0}\rightarrow \pm h$ and setting 
\begin{equation}
\varrho _{+}=\pi ^{+}(w;r_{0}=1,r_{1}=1,r_{2}=0)\quad \text{and\quad }%
\varrho _{-}=\pi ^{+}(w;r_{0}=1,r_{1}=0,r_{2}=1)\;.  \label{rho+-}
\end{equation}
Again we notice that the representation space in (\ref{pi2}) truncates and
it will be understood that the representation spaces associated with $%
\varrho _{\pm }$ are the modules obtained by acting freely with $e_{\pm }$
on the highest (lowest) weight vector $\left| 0\right\rangle $. Let $Q^{\pm
}=(\limfunc{Tr}_{\varrho \pm }\otimes 1)\mathbf{Q}$ then we find for the
eigenvalues associated with a Bethe vector (\ref{Bethev}), 
\begin{equation}
Q^{+}(z)=(-1)^{M}P_{B}(z)\sum_{\ell =0}^{\infty }\lambda ^{2\ell }q^{-2\ell
S^{z}}=\frac{(-1)^{M}P_{B}(z)}{1-\lambda ^{2}q^{-2S^{z}}}\;,  \label{Q+}
\end{equation}
and 
\begin{equation}
Q^{-}(z)=P_{B}(z)\sum_{\ell =0}^{\infty }\lambda ^{2\ell }q^{-2\ell S^{z}}%
\frac{\prod_{m=1}^{M}(zq^{2\ell +2}/\zeta _{m}-1)}{P_{B}(zq^{2\ell
+2})P_{B}(zq^{2\ell })}\;.  \label{Q-}
\end{equation}
For absolute convergence we have assumed that (\ref{qconv}) holds as
befrore. Notice that in order to satisfy the functional equation (\ref{TQ6v2}%
) with the transfer matrix we have to keep in mind that the operators $%
Q^{\pm }$ still depend implicitly on the parameters $(r_{0},r_{1},r_{2})$
which shift in (\ref{seq2}). This is particularly important in the case of $%
Q^{+}=Q_{\leq }(r_{0}=1,r_{1}=1,r_{2}=0)$ where in the functional equation
the shift $r_{1}\rightarrow r_{1}q^{\pm 2}$ occurs and, hence, the
truncation condition (\ref{truncc}) for the auxiliary space changes.

This highlights the importance of introducing the free parameters $%
(r_{0},r_{1},r_{2})$ in (\ref{pi2}). Even if one would directly start the
discussion with the simpler looking representations (\ref{rho+-}) the
decomposition of the tensor product (\ref{seq2}), which underlies the $TQ$%
-equation (\ref{TQ2}), leads to the consideration of the more general
representation (\ref{pi2}). (See also the comments in the introduction of 
\cite{KQ} for the root of unity case.) We should therefore understand the
expressions (\ref{Q+}) and (\ref{Q-}) as a decomposition of the general
solution (\ref{Qtrunc}). Namely, we can write (\ref{Qtrunc}) as a product of
(\ref{Q+}) and (\ref{Q-}) (compare with formula (4.10) in \cite{BLZ99}), 
\begin{equation}
Q_{\leq }(z;r_{0},r_{1})=(-1)^{M}r_{0}^{-S^{z}}(1-\lambda
^{2}q^{-2S^{z}})Q^{+}(zr_{1})Q^{-}(z)\;.  \label{Qdecomp}
\end{equation}

One now easily derives the analogue of the functional relations reported in 
\cite{BLZ99} for the Coulomb gas formalism. For instance, we find for the
eigenvalues the following formula corresponding to the ``quantum Wronskian
condition'' (cf equation (4.3) in \cite{BLZ99}), 
\begin{equation}
Q^{+}(zq^{2})Q^{-}(z)-\lambda ^{2}q^{-2S^{z}}Q^{+}(z)Q^{-}(zq^{2})=\frac{%
\prod_{m=1}^{M}(1-zq^{2}/\zeta _{m})}{1-\lambda ^{2}q^{-2S^{z}}}\;.
\label{qW}
\end{equation}
This is a special case of a more general relation (see equation (\ref{Qfus})
below) involving the fusion matrices of the six-vertex model, which we
discuss next.

\section{The six-vertex fusion hierarchy}

In order to provide further support for our algebraic Bethe ansatz
computation of the eigenvalues of the auxiliary matrices, we now make
contact with what in the literature is known as the fusion hierarchy. We
follow a similar line of argument to that in \cite{FM8v2} for the
eight-vertex model. First, we will briefly review the representation
theoretic aspects in the construction of the fusion hierarchy which will be
the key to connecting fusion and auxiliary matrices at roots of unity. We
specialize at once to the XXZ spin-chain (\ref{XXZchain}).

We start by introducing the monodromy matrices associated with the fusion
matrices. Denote by $\pi _{z}^{(n)}:U_{q}(\widetilde{sl}_{2})\rightarrow 
\limfunc{End}\mathbb{C}^{n+1}$ the spin $n/2$ evaluation representation of
the quantum loop algebra, i.e. 
\begin{eqnarray}
\pi _{z}^{(n)}(e_{1})\left| m\right\rangle &=&[n-m+1]_{q}\left|
m-1\right\rangle ,\quad \pi _{z}^{(n)}(f_{0})=z^{-1}\pi _{z}^{(n)}(e_{1}), 
\notag \\
\pi _{z}^{(n)}(f_{1})\left| m\right\rangle &=&[m+1]_{q}\left|
m+1\right\rangle ,\quad \pi _{z}^{(n)}(e_{0})=z\pi _{z}^{(n)}(f_{1}),  \notag
\\
\pi _{z}^{(n)}(q^{h_{1}})\left| m\right\rangle &=&q^{n-2m}\left|
m\right\rangle ,\quad \pi _{z}^{(n)}(q^{h_{0}})=\pi _{z}^{(n)}(q^{-h_{1}}),
\label{pin}
\end{eqnarray}
with $m=0,1,...,n$. Let 
\begin{eqnarray}
\mathbf{T}^{(n+1)}(zq^{-n-1}) &=&(\pi _{z}^{(n)}\otimes \pi _{\mathcal{H}%
})\lambda ^{h\otimes \mathbf{1}}\mathbf{R}\in \limfunc{End}\mathbb{C}%
^{n+1}\otimes \mathcal{H,}  \label{fusmom} \\
&=&\lambda ^{\pi ^{(n)}(h)\otimes 1}L_{0M}^{(n+1)}(z/\zeta _{M})\cdots
L_{01}^{(n+1)}(z/\zeta _{1}),\quad \quad \mathcal{H}=(\mathbb{C}%
^{2})^{\otimes M},  \notag
\end{eqnarray}
with 
\begin{equation}
L^{(n+1)}(w)=\left( \pi _{w}^{(n)}\otimes \pi _{z=1}^{(1)}\right) \mathbf{R}%
\in \limfunc{End}\mathbb{C}^{n+1}\otimes \mathbb{C}^{2}\;.  \label{fusL}
\end{equation}
Note that we have labelled the fusion matrices by the dimension of their
auxiliary space instead of the spin. The matrices (\ref{mom6v}), (\ref
{fusmom}) and (\ref{fusL}) again satisfy the Yang-Baxter relation. The
matrix elements of (\ref{fusL}) w.r.t. the representation $\pi _{z=1}^{(1)}$
are explicitly given by 
\begin{eqnarray}
\left\langle 0|L^{(n+1)}(w)|0\right\rangle &=&\rho _{+}\,\pi ^{(n)}(q^{\frac{%
h}{2}})-\rho _{-}\,\pi ^{(n)}(q^{-\frac{h}{2}}),\quad  \notag \\
\left\langle 0|L^{(n+1)}(w)|1\right\rangle &=&\rho _{+}\,(q-q^{-1})\pi
^{(n)}(q^{\frac{h}{2}})\pi ^{(n)}(f_{1}),  \notag \\
\left\langle 1|L^{(n+1)}(w)|0\right\rangle &=&\rho _{-}\,(q-q^{-1})\pi
^{(n)}(e_{1})\pi ^{(n)}(q^{-\frac{h}{2}}),\quad  \notag \\
\left\langle 1|L^{(n+1)}(w)|1\right\rangle &=&\rho _{+}\,\pi ^{(n)}(q^{-%
\frac{h}{2}})-\rho _{-}\,\pi ^{(n)}(q^{\frac{h}{2}})  \label{fusL2}
\end{eqnarray}
with the coefficients $\rho _{\pm }$ satisfying the constraint $\rho
_{+}/\rho _{-}=wq$. From the following non-split exact sequence describing
the decomposition of evaluation representations of the quantum loop algebra $%
U_{q}(\widetilde{sl}_{2})$ \cite{CPbook}, 
\begin{equation}
0\rightarrow \pi _{w^{\prime }}^{(n-1)}\overset{\imath }{\hookrightarrow }%
\pi _{w}^{(n)}\otimes \pi _{z}^{(1)}\overset{\tau }{\rightarrow }\pi
_{w^{\prime \prime }}^{(n+1)}\rightarrow 0,\quad w=w^{\prime
}q^{-1}=w^{\prime \prime }q=zq^{n+1}  \label{fusseq}
\end{equation}
one derives a functional relation, known as the fusion hierarchy (see e.g. 
\cite{KS,KR,RW02} and also \cite{Kun&Co} for an alternative form of the
fusion equation), 
\begin{equation}
T^{(n)}(z)T^{(2)}(zq^{-2})=T^{(n+1)}(zq^{-2})\prod_{m=1}^{M}\left(
zq^{2}/\zeta _{m}-1\right) +T^{(n-1)}(zq^{2})\prod_{m=1}^{M}\left( z/\zeta
_{m}-1\right) \;.  \label{fus}
\end{equation}
Here 
\begin{equation}
T^{(n+1)}(zq^{-n-1})=(\limfunc{Tr}_{\pi _{z}^{(n)}}\otimes 1)\mathbf{T}%
^{(n+1)}(zq^{-n-1})\in \limfunc{End}\mathcal{H}\;.  \label{fusmatrix}
\end{equation}
The coefficients in (\ref{fusL2}) have been chosen as 
\begin{equation}
\rho _{+}=wq,\quad \rho _{-}=1,  \label{rho}
\end{equation}
and we have identified 
\begin{equation}
T^{(2)}(zq^{-2})\equiv q^{-\frac{M}{2}}T(z)\prod_{m=1}^{M}(zq^{2}/\zeta
_{m}-1),\quad T^{(1)}(z)\equiv \prod_{m=1}^{M}(zq^{2}/\zeta _{m}-1)\;.
\end{equation}
The coefficient functions in (\ref{fus}) have been calculated using the
explicit form of the inclusion and projection map in (\ref{fusseq}), 
\begin{equation*}
\pi ^{(n-1)}\ni \left| m\right\rangle \overset{\imath }{\hookrightarrow }%
\left| m\right\rangle ^{\prime }=[n-m]\,\left| m\right\rangle \otimes \left|
1\right\rangle -q^{n-m}[m+1]\,\left| m+1\right\rangle \otimes \left|
0\right\rangle ,
\end{equation*}
and 
\begin{equation*}
\left| m\right\rangle ^{\prime \prime }=\frac{[n+1]}{[n-m+1]}\,\left|
m\right\rangle \otimes \left| 0\right\rangle \overset{\tau }{\rightarrow }%
\left| m\right\rangle \in \pi ^{(n+1)}\;.
\end{equation*}

Inserting the explicit expression for the eigenvalues of the six-vertex
transfer matrix from the algebraic Bethe ansatz (\ref{TABA}) we obtain the
following formula for the eigenvalues of the fusion matrices associated with
the Bethe states (\ref{Bethev}), 
\begin{equation}
T^{(n)}(z)=\lambda ^{-n-1}q^{(n+1)S^{z}}P_{B}(z)P_{B}(zq^{2n})\sum_{\ell
=1}^{n}\lambda ^{2\ell }q^{-2\ell S^{z}}\frac{\prod_{m=1}^{M}(zq^{2\ell
}/\zeta _{m}-1)}{P_{B}(zq^{2\ell })P_{B}(zq^{2\ell -2})}\;.  \label{fust}
\end{equation}
Here we have again denoted eigenvalues and operators by the same symbol. $%
P_{B}$ denotes the Bethe polynomial defined in (\ref{PB}) whose zeroes are
solutions to the Bethe ansatz equations (\ref{genBAE}). The proof follows
via induction and is straightforward. Notice that the eigenvalues (\ref{fust}%
) are polynomials in $z$. The rational function in (\ref{fust}) only has
poles at $z=z_{j},z_{j}q^{2n}$ which are cancelled by the factor $%
P_{B}(z)P_{B}(zq^{2n})$ in front of the sum. The residue of the remaining
poles is zero due to the Bethe ansatz equations.

Away from a root of unity we can already connect the fusion hierarchy with
the auxiliary matrices (\ref{Q+}) and (\ref{Q-}). A straightforward
calculation proves for the eigenvalues the relation 
\begin{equation}
\lambda ^{-n}q^{nS^{z}}Q^{+}(zq^{2n})Q^{-}(z)-\lambda
^{n}q^{-nS^{z}}Q^{+}(z)Q^{-}(zq^{2n})=\frac{(-1)^{M}\lambda ^{-2}q^{2S^{z}}}{%
\lambda ^{-1}q^{S^{z}}-\lambda q^{-S^{z}}}\,T^{(n)}(z)\;,  \label{Qfus}
\end{equation}
which corresponds to formula (4.1) in \cite{BLZ99}. The above functional
equation not only holds for the eigenvalues but also for the operators
provided the Bethe states constitute a complete basis of the quantum space.

\subsection{Degeneracies at roots of unity}

The discussion of the fusion matrices has so far applied to the case of
``generic'' $q$ (i.e. the deformation parameter being either a root of unity
or not) and finite solutions to the Bethe ansatz equations entering the
eigenvalues via the Bethe polynomial (\ref{PB}). We now specialize to the
case $q^{N^{\prime }}=\pm 1$ and start by showing that the fusion matrices (%
\ref{fusmatrix}) exhibit the same infinite-dimensional non-abelian
symmetries as the transfer matrix.

Recall from \cite{DFM,Ktw} that the symmetry generators for the XXZ
spin-chain are given by ($i=0,1$) 
\begin{eqnarray}
E_{i}^{(N^{\prime })} &=&\lim_{q^{\prime }\rightarrow
q}\bigotimes_{m=1}^{M}\pi _{\zeta _{m}}^{(1)}\,\Delta
^{(M)}(e_{i}^{N^{\prime }})/[N^{\prime }]_{q^{\prime }}!\,,  \label{E&F} \\
F_{i}^{(N^{\prime })} &=&\lim_{q^{\prime }\rightarrow
q}\bigotimes_{m=1}^{M}\pi _{\zeta _{m}}^{(1)}\,\Delta
^{(M)}(f_{i}^{N^{\prime }})/[N^{\prime }]_{q^{\prime }}!\;.
\end{eqnarray}
Here $\Delta ^{(M)}=(1\otimes \Delta )\Delta ^{(M-1)}$ denotes the $M$-fold
coproduct with $\Delta ^{(2)}\equiv \Delta $ and $q^{\prime }$ is some
number with $q^{\prime N^{\prime }}\neq \pm 1$. The above form of the
symmetry generators of $U(\widetilde{sl}_{2})$ is restricted to the
commensurate spin-sectors $2S^{z}=0\func{mod}N$ when $\lambda =1$ \cite{DFM}
but extends for the subalgebras $U(b_{\mp })$ to all spin-sectors when $%
\lambda =q^{\pm S^{z}}$ \cite{Ktw}. The extension of the loop symmetry from
the fusion degree $n=2$ to arbitrary $n$ \thinspace now simply follows by
induction from the recursion relation (\ref{fus}). Thus, the eigenspaces of
the fusion matrices also organize into multiplets containing states whose
total spin differs by multiples of $N^{\prime }$.\smallskip

\textbf{Remark}. In a similar manner one proves the loop symmetry of the
higher-spin six-vertex models. The only major difference lies in the choice
for the representation $\pi _{\mathcal{H}}$ defining the quantum space which
changes to $\pi _{\mathcal{H}}=\bigotimes_{m=1}^{M}\pi _{\zeta _{m}}^{(n)}$.
An explicit calculation shows that the analogue of the operators (\ref{E&F})
is well-defined. Adopting the proof in \cite{Ktw} for the loop symmetry of
the ``fundamental'' fusion matrix $T^{(2)}$ one shows again by induction
that $T^{(n+1)}$ exhibits the same symmetries. This provides an alternative
proof to the one given in \cite{KM01}.

\subsection{Spectrum of the fusion matrices at $q^{N}=1$}

In order to elucidate the spectrum of the fusion matrices at roots of unity
let us discuss the limit $q^{\prime }\rightarrow q$ from transcendental or
irrational $q^{\prime }$ to $q$ being a root of unity. In order to simplify
the notation we will often denote this limit by the symbol $q^{N}\rightarrow
1$ in the following. In order to perform the root of unity limit we need the
explicit dependence of the Bethe roots $z_{i}$ on the arbitrary deformation
parameter $q^{\prime }$. Unfortunately, this dependence is in general not
known. However, for small chains up to the size $M\leq 8$ one can compute
some Bethe roots analytically. According to these examples the following
scenarios might occur in the root of unity limit:

\begin{enumerate}
\item  A Bethe root tends to zero, 
\begin{equation}
\lim_{q^{N}\rightarrow 1}z_{i}=0\;.
\end{equation}
We shall denote the number of such Bethe roots by $n_{0}$.

\item  A Bethe root tends to infinity, 
\begin{equation}
\lim_{q^{N}\rightarrow 1}z_{i}=\infty \;.
\end{equation}
We shall denote the number of infinite roots by $n_{\infty }$.

\item  There are $N^{\prime }$ Bethe roots which form a complete string 
\begin{equation}
\lim_{q^{N}\rightarrow 1}(z_{i_{0}},z_{i_{1}},\ldots ,z_{i_{N^{\prime
}-1}})=(z_{i_{0}},z_{i_{1}}=z_{i_{0}}q^{2},\ldots ,z_{i_{N^{\prime
}-1}}=z_{i_{0}}q^{2N^{\prime }-2})\;.
\end{equation}
Note that these complete strings obtained in the root of unity limit do not
coincide with the zeroes of the classical Drinfeld polynomial (\ref{PS})
contained in the spectrum of the auxiliary matrices.
\end{enumerate}

Because of these three possibilities we deduce that taking the root of unity
limit of the Bethe polynomial (\ref{PB}) is not the same as the Bethe
polynomial containing the finite solutions of the Bethe ansatz equations 
\emph{at} a root of unity, i.e. 
\begin{equation*}
\lim_{q^{\prime }\rightarrow q}P_{B}(z,q^{\prime })\neq P_{B}(z,q)\quad 
\text{with\quad }q^{N^{\prime }}=\pm 1\;.
\end{equation*}
In the presence of vanishing Bethe roots the root of unity limit of the
Bethe polynomial might not even be well-defined. However, the eigenvalues of
the fusion and transfer matrices only contain ratios of these polynomials.
From the simple identities 
\begin{equation}
\lim_{z_{i}\rightarrow 0}\frac{(z-z_{i})(zq^{2n}-z_{i})}{(zq^{2\ell
}-z_{i})(zq^{2\ell -2}-z_{i})}=q^{2n-4\ell +2},
\end{equation}
and 
\begin{equation}
\lim_{z_{i}\rightarrow \infty }\frac{(z-z_{i})(zq^{2n}-z_{i})}{(zq^{2\ell
}-z_{i})(zq^{2\ell -2}-z_{i})}=1
\end{equation}
we infer that the spectrum of the fusion matrices at a root unity changes to 
\begin{equation}
\lim_{q^{N}\rightarrow 1}T^{(n)}(z)=\lambda
^{-n-1}q^{(n+1)s}P_{B}(z)P_{B}(zq^{2n})\sum_{\ell =1}^{n}\lambda ^{2\ell
}q^{-2\ell s}\frac{\prod_{m=1}^{M}(zq^{2\ell }/\zeta _{m}-1)}{%
P_{B}(zq^{2\ell })P_{B}(zq^{2\ell -2})}\;.  \label{Tnq1}
\end{equation}
Here we have set 
\begin{equation}
s=2n_{0}+S^{z}\func{mod}N^{\prime }\;.  \label{s}
\end{equation}
Again we remind the reader that $P_{B}$ is now the ``reduced'' Bethe
polynomial, i.e. it only contains the \emph{finite} Bethe roots at $q^{N}=1$
and no complete strings. From the identity (\ref{Tnq1}) it follows by a
direct calculation that one has the following simplification of the fusion
hierarchy in terms of the eigenvalues 
\begin{equation}
\lim_{q^{N}\rightarrow 1}T^{(N^{\prime }+1)}(z)=(\lambda ^{-N^{\prime
}}q^{N^{\prime }s}+\lambda ^{N^{\prime }}q^{-N^{\prime
}s})\prod_{m=1}^{M}(zq^{2}/\zeta _{m}-1)+\lim_{q^{N}\rightarrow
1}T^{(N^{\prime }-1)}(zq^{2})\;.  \label{trunc}
\end{equation}
This formula has been reported for $\lambda =1,\;\zeta _{m}=1$ before in the
literature \cite{Kun2,Nepo,FM8v2}.

\subsection{Connection with auxiliary matrices at roots of 1}

We will now argue that the fusion matrix of degree $N^{\prime }$ at $%
q^{N^{\prime }}=\pm 1$ can be identified with a special limit of the
auxiliary matrix (\ref{T&Q6v}). First, we observe that the corresponding
auxiliary spaces have the same dimension. From the representation theory of
quantum groups at roots of unity \cite{CK,CKP} one now deduces the
following. Restrict the respective evaluation representations (\ref{pi1})
and (\ref{pin}) of the quantum loop algebra to the finite subalgebra $%
U_{q}(sl_{2})$. If the values of the central elements in the respective $%
U_{q}(sl_{2})$-representations determining the auxiliary spaces are equal
then they must be isomorphic. The values of the central elements at a root
of unity are 
\begin{eqnarray*}
T^{(N^{\prime })} &:&\quad \quad \pi ^{(N^{\prime })}(e_{1}^{N^{\prime
}})=\pi ^{(N^{\prime })}(f_{1}^{N^{\prime }})=0,\text{\quad }\pi
^{(N^{\prime })}(q^{h})^{N^{\prime }}=q^{N^{\prime }(N^{\prime }-1)} \\
Q_{\mu } &:&\quad \quad \pi ^{\mu }(e_{1}^{N^{\prime }})=\pi ^{\mu
}(f_{1}^{N^{\prime }})=0,\text{\quad }\pi ^{\mu }(q^{h})^{N^{\prime
}}=q^{-N^{\prime }}\mu ^{-N^{\prime }}
\end{eqnarray*}
and for the Casimir $c=q^{h+1}+q^{-h-1}+(q-q^{-1})^{2}fe$ one finds 
\begin{eqnarray*}
T^{(N^{\prime })} &:&\quad \quad \pi ^{(N^{\prime })}(c)=q^{N^{\prime
}}+q^{-N^{\prime }},\quad  \\
Q_{\mu } &:&\quad \quad \pi ^{\mu }(c)=\mu +\mu ^{-1}\;.
\end{eqnarray*}
Hence, in the limit $\mu \rightarrow q^{N^{\prime }}=\pm 1$ both
representation become isomorphic and one can therefore construct a gauge
transformation in the auxiliary space rendering the fusion and auxiliary
matrix equal (see \cite{KQ} for details). Thus, we conclude 
\begin{equation}
\lim_{q^{N}\rightarrow 1}T^{(N^{\prime })}(z)=\lim_{\mu \rightarrow
q^{N^{\prime }}}Q_{\mu }(z/\mu )\;.  \label{TNQ}
\end{equation}
Note that in this limit the auxiliary matrix therefore also becomes
degenerate. Nevertheless, we can check the eigenvalues for consistency be
comparing the expressions for the eigenvalues derived from the algebraic
Bethe ansatz with the ones from the fusion hierarchy. One finds for the
eigenvalues of the fusion matrix 
\begin{equation}
\lim_{q^{N}\rightarrow 1}T^{(N^{\prime })}(z)=\lambda ^{-N^{\prime
}-1}q^{(N^{\prime }+1)s}P_{B}(z)P_{B}(zq^{2N^{\prime }})\sum_{\ell \in 
\mathbb{Z}_{N^{\prime }}}\frac{\lambda ^{2\ell }q^{-2\ell
s}\prod_{m=1}^{M}(zq^{2\ell }/\zeta _{m}-1)}{P_{B}(zq^{2\ell
})P_{B}(zq^{2\ell -2})}\;.  \label{fusTN}
\end{equation}
In the commensurate sectors $2S^{z}=0\func{mod}N$ where $n_{0}=n_{\infty }=0$
we can set $s=0$ in (\ref{fusTN}). Then the spectrum of the auxiliary
matrices $Q_{\mu }$ derived for finite Bethe states at a root of unity
coincides for $\mu =q^{N^{\prime }}=\pm 1$ with the one of the fusion
matrix. Outside the commensurate sectors it has been argued (and proven for $%
N=3$) in \cite{KQ2} that the auxiliary matrices (\ref{T&Q6v}) have
eigenvalues of the form (\ref{19}). When comparing with formula (11) in \cite
{KQ2} one has to make the replacement $\bar{n}_{\infty }\rightarrow
n_{\infty }$ in the notation. Here $P_{S}$ denotes the previously discussed
polynomial (\ref{PS}) whose zeroes are fixed by the representation theory of
the loop algebra up to an factor $\mu ^{2N^{\prime }}$, 
\begin{equation}
q^{N^{\prime }}=1:\quad \mathcal{N}_{\mu }\,z^{\bar{n}_{\infty
}}P_{S}(z^{N^{\prime }},\mu =q^{N^{\prime }})=q^{(N^{\prime }+1)s}\sum_{\ell
\in \mathbb{Z}_{N^{\prime }}}\frac{q^{-2\ell s}\prod_{m=1}^{M}(zq^{2\ell
}/\zeta _{m}-1)}{P_{B}(zq^{2\ell })P_{B}(zq^{2\ell -2})}\;.
\end{equation}
The power of the monomial must be an integer by construction of the
auxiliary matrix and the normalization chosen in (\ref{G1}). Interestingly,
numerical examples show that vanishing and infinite Bethe roots seem to be
absent for all spin-sectors when $\lambda =q^{\pm S^{z}}$, i.e. for the
boundary conditions where the symmetry generators of the loop algebra are
known for all spin-sectors \cite{Ktw}. There appears to be a close
connection between the form of the symmetry algebra generators and the
presence of infinite or vanishing Bethe roots. This might hold the clue for
constructing the symmetry algebra outside the commensurate sectors when $%
\lambda =1$. Further investigation is needed to clarify this point.

\section{Conclusions}

In this article we have connected two known alternative methods of solving
integrable models, the algebraic Bethe ansatz and Baxter's concept of
auxiliary matrices. Within the community of integrable models the notion of
auxiliary matrices has historically received less attention than the Bethe
ansatz, even though it applies to a wider range of models. Unlike the Bethe
ansatz auxiliary matrix do not rely on the existence of a pseudo-vacuum or
spin conservation; see the discussion in Section 3. One possible reason that
auxiliary matrices have not earlier been investigated in more detail, might
be that Baxter's construction procedure \cite{BxBook} leads to a final
auxiliary matrix whose algebraic structure is not of the simple form (\ref
{T&Q6v}). Notice also that Baxter's auxiliary matrix for $\lambda =1$ is
limited to the case of even spin-chains, $M\in 2\mathbb{N}$, when $q$ is not
a root of unity and to $M\func{mod}N\in 2\mathbb{N}$ when $q^{N}=1$; see
condition (9.8.16) in \cite{BxBook}. For a construction of Baxter's
auxiliary matrices at more general boundary conditions see e.g. \cite{YB95}.

In contrast the construction of auxiliary matrices using intertwiners of
quantum groups yields simpler algebraic expressions, has no restrictions on
the length of the spin-chain and, most importantly, allows for maintaining
the Yang-Baxter relation (\ref{YBE}). It is the last fact which put us in
the position to link the formalism of the algebraic Bethe ansatz with the $Q$%
-operator or auxiliary matrix. Furthermore, the use of representation theory
in the construction of $Q$-operators enables one to make direct contact with
the underlying algebraic structure of the integrable model, the quantum loop
algebra. The auxiliary matrices discussed in the present article are
therefore a key link in understanding the relation between the Bethe ansatz
and the representation theory of quantum groups.\smallskip

The importance of this aspect of the $Q$-matrix is particularly highlighted
in the root of unity case. As pointed out in \cite{KQ2} the auxiliary
matrices constructed at a root of unity \cite{KQ} might serve as an
efficient tool to analyze the irreducible representations of the loop
algebra symmetry. In our investigation of the spectrum of the auxiliary
matrices via the Bethe ansatz as well as the fusion hierarchy we came across
the polynomial (cf formula (\ref{PS}) in the text) 
\begin{equation}
\mathcal{N}_{\mu =1}P_{S}(z^{N^{\prime }},\mu =1)=\sum_{k\in \mathbb{Z}%
_{N^{\prime }}}\frac{\lambda ^{-2k}q^{2kS^{z}}(zq^{-2k}-1)^{M}}{%
P_{B}(zq^{-2k})P_{B}(zq^{-2k-2})}\,,\quad q^{N^{\prime }}=\pm 1\;.
\label{PS2}
\end{equation}
Recall that the Bethe ansatz equations are sufficient to ensure that the
right hand side of this identity is a polynomial. This expression has been
conjectured to be the classical Drinfeld polynomial \cite{FM01b}. The latter
describes the irreducible representation of the loop algebra spanning the
degenerate eigenspace of the transfer matrix; see \cite{KQ2} for concrete
examples.\smallskip

What is the benefit of the identification? The symmetry algebra can be used
to compute the Drinfeld polynomial, i.e. the left hand side of equation (\ref
{PS2}). We can then use the above identity to extract the Bethe roots for
each degenerate eigenspace of the transfer matrix. At the moment this
identification still awaits proof as well as its extension to the
quasi-periodic boundary conditions $\lambda =q^{\pm S^{z}}$ investigated in 
\cite{Ktw}. The results of this paper provide a further step towards this
direction.\smallskip

Away from a root of unity our interest has been to make contact with the
findings in \cite{BLZ97,BLZ99} for the Coulomb gas formalism of conformal
field theory. This also provided the motivation to perform the algebraic
Bethe ansatz computation of the spectrum of the $Q$-operators for an
arbitrary quantum space in order to accommodate the CFT setup. For the XXZ
spin-chain we started from the results obtained in \cite{RW02} to analyze
the spectrum of the auxiliary matrix and to find the analogue of the
functional equations reported in \cite{BLZ99} for the six-vertex model. The
noteworthy difference between the constructions in \cite{BLZ99} and \cite
{RW02} is the occurrence of free parameters in the representation spanning
the auxiliary space of the $Q$-operator; see definition (\ref{pi2}). We saw
that the introduction of such parameters is natural in light of the
representation theoretic background (\ref{seq2}) of the $TQ$-equation (\ref
{TQ2}); see also \cite{RW02}.\smallskip 

What is the role of these parameters with hindsight of the spectrum of the
six-vertex model? Our result for the eigenvalues (\ref{Qtrunc}) and (\ref{RQ}%
) showed that the parameter $r_{0}$ in (\ref{pi2}) is needed to break
spin-reversal symmetry, while the parameters $r_{1,2}$ simply reflect the
polynomial structure of the eigenvalues of the auxiliary matrix in the
spectral variable $z$. One of them is needed to truncate the auxiliary space
in order to achieve convergence when $q$ is not a root of unity. The freedom
in choosing the remaining parameter can be used to decompose the eigenvalue (%
\ref{Qtrunc}) into the polynomials (\ref{Q+}) and (\ref{Q-}); see also (\ref
{Qdecomp}). This decomposition has been observed in \cite{BLZ99} in the
context of CFT (see equation (4.10) therein) and was the starting point for
the formulation of a series of functional equations all of which we
recovered in the case of the XXZ spin-chain; see for example equation (\ref
{Qfus}) in the text. As these functional relations have been the starting
point for connecting the spectrum of the auxiliary matrices with ordinary
differential equations \cite{ODEIM}, it is natural to ask whether this can
be also achieved for the XXZ spin-chain. So far partial results exist at
some roots of unity only \cite{ODEXXZ}.\smallskip 

Convergence problems did not arise in the root-of-unity case and one might
ask about the root of unity limit of the auxiliary matrix (\ref{Qtrunc}).
While we did not pursue this issue in detail it seems at first sight
plausible that in this limit the infinite-dimensional representation splits
up into an infinite number of finite-dimensional subrepresentations.
Reducing the trace of the monodromy matrix (\ref{momQ}) to such a
subrepresentation one might be lead to the conclusion that one ends up with
the auxiliary matrices constructed \emph{at} a root of unity in \cite{KQ}.
This is not true. The full set of auxiliary matrices constructed in \cite{KQ}
does not preserve the total spin $S^{z}$. Here we only dealt with a subset
of the auxiliary matrices available at a root of unity. These additional $Q$%
-operators also contain free parameters but their nature is different from
the ones in (\ref{pi2}). They reflect the enhanced symmetry of the XXZ
spin-chain at rational coupling, i.e. $q^{N^{\prime }}=\pm 1$. The full
scope of this symmetry is not accessible via the Bethe ansatz whence further
work is required to determine the spectrum of all auxiliary matrices in \cite
{KQ} as well as the spectrum inside the degenerate eigenspaces; see our
discussion in Section 4.1.2.{\small \medskip }

\noindent \textbf{Acknowledgments}. It is a pleasure to thank Harry Braden,
Barry McCoy, Marco Rossi and Robert Weston for discussions. This work has
been financially supported by the EPSRC Grant GR/R93773/01.

\appendix

\section{The spectrum of $Q$. Proof for $n_{B}=2,3$.}

We outline the main steps in proving the conjecture (\ref{Con}) for the
Bethe states (\ref{Bethev}) with $n_{B}=2,3$. We will make use of the
following well-known commutation relation of the Yang-Baxter algebra, 
\begin{eqnarray*}
A_{1}B_{2} &=&\frac{1}{b_{21}}\,B_{2}A_{1}-\frac{c_{21}}{b_{21}}%
\,B_{1}A_{2},\quad D_{1}B_{2}=\frac{1}{b_{12}}\,B_{2}D_{1}-\frac{%
c_{12}^{\prime }}{b_{12}}\,B_{1}D_{2}, \\
\lbrack D_{1},A_{2}] &=&\frac{c_{12}}{b_{12}}B_{2}C_{1}-\frac{c_{12}^{\prime
}}{b_{12}}B_{1}C_{2},\quad \lbrack C_{1},B_{2}]=\frac{c_{12}^{\prime }}{%
b_{12}}\,\left( A_{2}D_{1}-A_{1}D_{2}\right) \;.
\end{eqnarray*}
Here the indices are a shorthand notation for the dependence of the
operators (\ref{ABCD}) and the Boltzmann weights (\ref{abc}) on the
respective Bethe roots. For instance, 
\begin{equation*}
b_{ij}\equiv b(z_{i}/z_{j}),\;...\quad \text{and\quad }A_{i}\equiv
A(z_{i}),\;B_{i}\equiv B(z_{i}),\;...\quad \text{etc.}
\end{equation*}
To unburden the formulas we also introduce the symbols 
\begin{equation*}
\Lambda _{kl}^{i}\equiv \frac{\delta _{l}(w/z_{i})}{\alpha _{k}(w/z_{i})}-%
\frac{\beta _{l+1}(w/z_{i})\gamma _{l}(w/z_{i})}{\alpha
_{l+1}(w/z_{i})\alpha _{k}(w/z_{i})}\quad \quad \text{and\quad \quad }%
r_{k}^{i}\equiv \frac{\beta _{k}(w/z_{i})}{\alpha _{k}(w/z_{i})}\;.
\end{equation*}

\paragraph{The case $n_{B}=2$.}

From the relations (\ref{QA}), (\ref{QB}), (\ref{QC}) and (\ref{QD}) one
obtains 
\begin{eqnarray*}
Q_{kk}B_{1}B_{2} &=&\Lambda _{kk}^{1}\Lambda _{kk}^{2}B_{1}B_{2}Q_{kk} \\
&&+\frac{r_{k+1}^{1}}{b_{21}}\,Q_{kk+1}B_{2}A_{1}+\Lambda
_{kk}^{1}r_{k+1}^{2}\,B_{1}Q_{kk+1}A_{2}-r_{k+1}^{1}\frac{c_{21}}{b_{21}}%
\,Q_{kk+1}B_{1}A_{2} \\
&&-\frac{r_{k}^{1}}{b_{12}}\,Q_{k-1\,k}B_{2}D_{1}-\Lambda
_{kk}^{1}r_{k}^{2}\,B_{1}Q_{k-1\,k}D_{2}+r_{k}^{1}\frac{c_{12}^{\prime }}{%
b_{12}}\,Q_{k-1k}B_{1}D_{2} \\
&&+r_{k+1}^{1}r_{k}^{1}\,Q_{k-1\,k+1}B_{2}C_{1}+\Lambda
_{kk}^{1}r_{k+1}^{2}r_{k}^{2}\,B_{1}Q_{k-1\,k+1}C_{2} \\
&&+r_{k+1}^{1}r_{k}^{1}\frac{c_{12}^{\prime }}{b_{12}}\,Q_{k-1%
\,k+1}(A_{2}D_{1}-A_{1}D_{2})
\end{eqnarray*}
Moving all the $B$-operators past the $Q$-operators except for the first
term, this can be rewritten as 
\begin{multline*}
Q_{kk}B_{1}B_{2}=\Lambda _{kk}^{1}\Lambda _{kk}^{2}B_{1}B_{2}Q_{kk}+\tfrac{%
r_{k+1}^{1}}{b_{21}}\,Q_{kk+1}B_{2}A_{1}+\tfrac{r_{k+1}^{2}}{b_{12}}%
\,Q_{kk+1}B_{1}A_{2} \\
-\tfrac{r_{k}^{1}}{b_{12}}\,Q_{k-1\,k}B_{2}D_{1}-\tfrac{r_{k}^{2}}{b_{21}}%
\,Q_{k-1k}B_{1}D_{2}-r_{k+2}^{1}r_{k+1}^{2}\tfrac{\Lambda _{kk}^{1}}{\Lambda
_{kk+1}^{1}}\,Q_{kk+2}A_{1}A_{2} \\
-r_{k}^{2}r_{k-1}^{1}\tfrac{\Lambda _{kk}^{1}}{\Lambda _{k-1\,k}^{1}}%
\,Q_{k-2k}D_{1}D_{2}+\tfrac{r_{k}^{1}r_{k+1}^{2}}{b_{12}}%
\,Q_{k-1k+1}A_{2}D_{1}+\tfrac{r_{k+1}^{1}r_{k}^{2}}{b_{21}}%
\,Q_{k-1\,k+1}A_{1}D_{2} \\
+r_{k}^{1}r_{k+1}^{1}\tfrac{\Lambda _{kk}^{2}}{\Lambda _{kk+1}^{2}b_{12}}%
\,Q_{k-1\,k+1}B_{2}C_{1}+r_{k}^{2}r_{k+1}^{2}\tfrac{\Lambda _{kk}^{1}}{%
\Lambda _{kk+1}^{1}b_{21}}\,Q_{k-1\,k+1}B_{1}C_{2} \\
-r_{k+2}^{1}r_{k+1}^{2}r_{k}^{1}\tfrac{\Lambda _{kk}^{1}}{\Lambda _{kk+1}^{1}%
}\,Q_{k-1k+2}C_{1}A_{2}-r_{k+2}^{1}r_{k+1}^{2}r_{k}^{2}\tfrac{\Lambda
_{kk}^{1}}{\Lambda _{k-1\,k+1}^{1}}\,Q_{k-1k+2}A_{1}C_{2} \\
+r_{k+1}^{1}r_{k}^{2}r_{k-1}^{1}\tfrac{\Lambda _{kk}^{1}}{\Lambda
_{k-1\,k}^{1}}\,Q_{k-2k+1}C_{1}D_{2}+r_{k+1}^{2}r_{k}^{2}r_{k-1}^{1}\tfrac{%
\Lambda _{kk}^{1}}{\Lambda _{k-1\,k+1}^{1}}\,Q_{k-2k+1}D_{1}C_{2} \\
-r_{k+2}^{1}r_{k+1}^{2}r_{k}^{2}r_{k-1}^{1}\tfrac{\Lambda _{kk}^{1}}{\Lambda
_{k-1\,k+1}^{1}}\,Q_{k-2k+2}C_{1}C_{2}
\end{multline*}
Acting with this expression on the pseudovacuum all the terms containing $C$%
-operators vanish. Taking the trace on both sides of the equation we then
obtain 
\begin{multline*}
\sum_{k}Q_{kk}B_{1}B_{2}\left| 0\right\rangle _{\mathcal{H}}=\sum_{k}\Lambda
_{kk}^{1}\Lambda _{kk}^{2}B_{1}B_{2}Q_{kk}\left| 0\right\rangle _{\mathcal{H}%
} \\
+\sum_{k}r_{k+1}^{1}\,Q_{kk+1}B_{2}(A_{1}/b_{21}-D_{1}/b_{12})\left|
0\right\rangle _{\mathcal{H}} \\
+\sum_{k}\frac{r_{k+1}^{2}}{b_{12}}%
\,Q_{kk+1}B_{1}(A_{2}/b_{12}-D_{2}/b_{21})\left| 0\right\rangle _{\mathcal{H}%
} \\
+\sum_{k}\left( \frac{r_{k+1}^{1}r_{k}^{2}}{b_{21}}\,Q_{k-1\,k+1}A_{1}D_{2}+%
\frac{r_{k}^{1}r_{k+1}^{2}}{b_{12}}\,Q_{k-1k+1}A_{2}D_{1}\right) \left|
0\right\rangle _{\mathcal{H}} \\
-\sum_{k}\left( r_{k+2}^{1}r_{k+1}^{2}\frac{\Lambda _{kk}^{1}}{\Lambda
_{kk+1}^{1}}\,Q_{kk+2}A_{1}A_{2}+r_{k}^{2}r_{k-1}^{1}\frac{\Lambda _{kk}^{1}%
}{\Lambda _{k-1\,k}^{1}}\,Q_{k-2k}D_{2}D_{1}\right) \left| 0\right\rangle _{%
\mathcal{H}}
\end{multline*}
Employing the Bethe ansatz equations (\ref{genBAE}), 
\begin{equation*}
\left\langle 0|A_{i}|0\right\rangle _{\mathcal{H}}\,b_{ij}=b_{ji}\,\left%
\langle 0|D_{i}|0\right\rangle _{\mathcal{H}},\quad i,j=1,2,
\end{equation*}
and using the identity 
\begin{equation*}
\frac{r_{k+1}^{1}r_{k}^{2}}{b_{12}}+\frac{r_{k+1}^{2}r_{k}^{1}}{b_{21}}%
-r_{k+1}^{1}r_{k}^{2}\frac{\Lambda _{k-1\,k-1}^{1}}{\Lambda _{k-1\,k}^{1}}%
-r_{k+1}^{2}r_{k}^{1}\frac{\Lambda _{k+1\,k+1}^{1}}{\Lambda _{kk+1}^{1}}=0
\end{equation*}
we see that the remaining terms cancel. This proves (\ref{Con}) for $n_{B}=2$%
.

\paragraph{The case $n_{B}=3$.}

The calculation follows the same strategy as before. The various steps can
be simplified by observing that the expressions must be symmetric under any
permutation of the Bethe roots. We omit the details of the computation and
only provide some intermediate steps. Starting from the result for the case $%
n_{B}=2$ one has 
\begin{multline*}
Q_{kk}B_{1}B_{2}B_{3}\left| 0\right\rangle _{\mathcal{H}}=\Lambda
_{kk}^{1}\Lambda _{kk}^{2}B_{1}B_{2}Q_{kk}B_{3}\left| 0\right\rangle _{%
\mathcal{H}} \\
+Q_{kk+1}\left( \tfrac{r_{k+1}^{2}}{b_{32}b_{12}}B_{1}B_{3}A_{2}-\tfrac{%
r_{k+1}^{2}c_{32}}{b_{12}b_{32}}B_{1}B_{2}A_{3}+\tfrac{r_{k+1}^{1}}{%
b_{21}b_{31}}B_{2}B_{3}A_{1}-\tfrac{r_{k+1}^{1}c_{31}}{b_{21}b_{31}}%
B_{2}B_{1}A_{3}\right) \left| 0\right\rangle _{\mathcal{H}} \\
-Q_{k-1k}\left( \tfrac{r_{k}^{2}}{b_{21}b_{23}}B_{1}B_{3}D_{2}-\tfrac{%
r_{k}^{2}c_{23}^{\prime }}{b_{21}b_{23}}B_{1}B_{2}D_{3}+\tfrac{r_{k}^{1}}{%
b_{12}b_{13}}B_{2}B_{3}D_{1}-\tfrac{r_{k}^{1}c_{13}^{\prime }}{b_{12}b_{13}}%
B_{2}B_{1}D_{3}\right) \left| 0\right\rangle _{\mathcal{H}} \\
+Q_{k-1k+1}\tfrac{r_{k+1}^{1}r_{k}^{2}}{b_{21}}\left( \tfrac{1}{b_{31}b_{23}}%
B_{3}A_{1}D_{2}-\tfrac{c_{31}}{b_{31}b_{23}}B_{1}A_{3}D_{2}-\tfrac{%
c_{23}^{\prime }}{b_{21}b_{23}}B_{2}A_{1}D_{3}+\tfrac{c_{21}c_{23}^{\prime }%
}{b_{21}b_{23}}B_{1}A_{2}D_{3}\right) \left| 0\right\rangle _{\mathcal{H}} \\
+Q_{k-1k+1}\tfrac{r_{k+1}^{2}r_{k}^{1}}{b_{12}}\left( \tfrac{1}{b_{32}b_{13}}%
B_{3}A_{2}D_{1}-\tfrac{c_{32}}{b_{32}b_{13}}B_{2}A_{3}D_{1}-\tfrac{%
c_{13}^{\prime }}{b_{12}b_{13}}B_{1}A_{2}D_{3}+\tfrac{c_{12}c_{13}^{\prime }%
}{b_{12}b_{13}}B_{2}A_{1}D_{3}\right) \left| 0\right\rangle _{\mathcal{H}} \\
+Q_{k-1k+1}\left( \tfrac{\Lambda _{kk}^{1}c_{23}^{\prime }}{\Lambda
_{kk+1}^{1}b_{23}}r_{k+1}^{2}r_{k}^{2}B_{1}(A_{3}D_{2}-A_{2}D_{3})+\tfrac{%
\Lambda _{kk}^{2}c_{13}^{\prime }}{\Lambda _{kk+1}^{2}b_{13}}%
r_{k+1}^{1}r_{k}^{1}B_{2}(A_{3}D_{1}-A_{1}D_{3})\right) \left|
0\right\rangle _{\mathcal{H}} \\
-Q_{kk+2}\tfrac{\Lambda _{kk}^{1}r_{k+2}^{1}r_{k+1}^{2}}{\Lambda _{kk+1}^{1}}%
\left( \tfrac{1}{b_{31}b_{32}}B_{3}A_{1}A_{2}-\tfrac{c_{31}}{b_{31}b_{32}}%
B_{1}A_{3}A_{2}+\tfrac{c_{21}c_{32}}{b_{21}b_{32}}B_{1}A_{2}A_{3}-\tfrac{%
c_{32}}{b_{32}b_{21}}B_{2}A_{1}A_{3}\right) \left| 0\right\rangle _{\mathcal{%
H}} \\
-Q_{k-2k}\tfrac{\Lambda _{kk}^{1}r_{k-1}^{1}r_{k}^{2}}{\Lambda _{k-1k}^{1}}%
\left( \tfrac{1}{b_{13}b_{23}}B_{3}D_{1}D_{2}-\tfrac{c_{13}^{\prime }}{%
b_{13}b_{23}}B_{1}D_{3}D_{2}+\tfrac{c_{12}^{\prime }c_{23}^{\prime }}{%
b_{12}b_{23}}B_{1}D_{2}D_{3}-\tfrac{c_{23}^{\prime }}{b_{23}b_{12}}%
B_{2}D_{1}D_{3}\right) \left| 0\right\rangle _{\mathcal{H}} \\
-Q_{k-1k+2}\left( \tfrac{\Lambda
_{kk}^{2}r_{k}^{1}r_{k}^{2}r_{k+2}^{2}c_{23}^{\prime }}{\Lambda
_{kk+1}^{2}b_{21}b_{23}}A_{1}(A_{3}D_{2}-A_{2}D_{3})+\tfrac{\Lambda
_{kk}^{1}r_{k}^{2}r_{k}^{1}r_{k+2}^{1}c_{13}^{\prime }}{\Lambda
_{kk+1}^{1}b_{12}b_{13}}A_{2}(A_{3}D_{1}-A_{1}D_{3})\right) \left|
0\right\rangle _{\mathcal{H}} \\
+Q_{k-2k+1}\left( \tfrac{\Lambda
_{kk}^{2}r_{k-1}^{2}r_{k}^{1}r_{k+1}^{2}c_{23}^{\prime }}{\Lambda
_{k-1k}^{2}b_{12}b_{23}}D_{1}(A_{3}D_{2}-A_{2}D_{3})+\tfrac{\Lambda
_{kk}^{1}r_{k-1}^{1}r_{k}^{2}r_{k+1}^{1}c_{13}^{\prime }}{\Lambda
_{k-1k}^{1}b_{21}b_{13}}D_{2}(A_{3}D_{1}-A_{1}D_{3})\right) \left|
0\right\rangle _{\mathcal{H}}
\end{multline*}
Here we have dropped all terms where $C$ acts on the pseudovacuum first and
those which contain more $C$ than $B$-operators. The last identity is
further simplified to 
\begin{multline*}
Q_{kk}B_{1}B_{2}B_{3}\left| 0\right\rangle _{\mathcal{H}}=\Lambda
_{kk}^{1}\Lambda _{kk}^{2}\Lambda _{kk}^{3}B_{1}B_{2}B_{3}Q_{kk}\left|
0\right\rangle _{\mathcal{H}} \\
+Q_{kk+1}\left( \frac{r_{k+1}^{1}}{b_{21}b_{31}}B_{2}B_{3}A_{1}+\frac{%
r_{k+1}^{2}}{b_{12}b_{32}}B_{1}B_{3}A_{2}+\frac{r_{k+1}^{3}}{b_{13}b_{23}}%
B_{1}B_{2}A_{3}\right) \left| 0\right\rangle _{\mathcal{H}} \\
-Q_{k-1k}\left( \frac{r_{k}^{1}}{b_{13}b_{12}}B_{2}B_{3}D_{1}+\frac{r_{k}^{2}%
}{b_{21}b_{23}}B_{1}B_{3}D_{2}+\frac{r_{k}^{3}}{b_{31}b_{32}}%
B_{1}B_{2}D_{3}\right) \left| 0\right\rangle _{\mathcal{H}} \\
+Q_{k-1k+1}\left( \frac{r_{k+1}^{1}r_{k}^{2}}{b_{21}b_{31}b_{23}}%
B_{3}A_{1}D_{2}+\frac{r_{k}^{1}r_{k+1}^{2}}{b_{12}b_{32}b_{13}}%
B_{3}A_{2}D_{1}+...\right) \left| 0\right\rangle _{\mathcal{H}} \\
-Q_{kk+2}\left( \frac{\Lambda _{kk}^{1}r_{k+2}^{1}r_{k+1}^{2}}{\Lambda
_{kk+1}^{1}b_{31}b_{32}}B_{3}A_{1}A_{2}+...\right) \left| 0\right\rangle _{%
\mathcal{H}}-Q_{k-2k}\left( \frac{\Lambda _{kk}^{1}r_{k-1}^{1}r_{k}^{2}}{%
\Lambda _{k-1k}^{1}b_{13}b_{23}}B_{3}D_{1}D_{2}+...\right) \left|
0\right\rangle _{\mathcal{H}} \\
-Q_{k-1k+2}\left( \frac{\Lambda _{kk}^{1}r_{k}^{3}r_{k+1}^{2}r_{k+2}^{1}}{%
\Lambda _{kk+1}^{1}b_{31}b_{32}}A_{1}A_{2}D_{3}+...\right) \left|
0\right\rangle _{\mathcal{H}} \\
+Q_{k-2k+1}\left( \frac{\Lambda _{kk}^{1}r_{k-1}^{1}r_{k}^{2}r_{k+1}^{3}}{%
\Lambda _{k-1k}^{1}b_{13}b_{23}}D_{1}D_{2}A_{3}+...\right) \left|
0\right\rangle _{\mathcal{H}} \\
+\frac{\Lambda _{kk}^{1}\Lambda _{kk}^{2}}{\Lambda _{kk+1}^{1}\Lambda
_{kk+2}^{2}}r_{k+1}^{3}r_{k+2}^{1}r_{k+3}^{2}Q_{kk+3}A_{1}A_{2}A_{3}\left|
0\right\rangle _{\mathcal{H}}-\frac{\Lambda _{kk}^{1}\Lambda _{kk}^{2}}{%
\Lambda _{k-1k}^{1}\Lambda _{k-2k}^{2}}%
r_{k}^{3}r_{k-1}^{1}r_{k-2}^{2}Q_{k-3k}D_{1}D_{2}D_{3}\left| 0\right\rangle
_{\mathcal{H}}\;.
\end{multline*}
The omitted terms in the parentheses are obtained by symmetrization w.r.t.
the Bethe roots. Again one computes that upon taking the trace on both sides
of the equation and invoking the Bethe ansatz equations, 
\begin{equation*}
\left\langle 0|A_{i}|0\right\rangle _{\mathcal{H}}\prod_{j\neq
i}b_{ij}=\left\langle 0|D_{i}|0\right\rangle _{\mathcal{H}}\prod_{j\neq
i}b_{ji},
\end{equation*}
all terms vanish except the first one. This proves (\ref{Con}) for $n_{B}=3$%
. We leave the case for general $n_{B}$ to a future calculation. Here we
shall be content with supporting the conjecture for $n_{B}>3$ by making
contact with various functional equations for the eigenvalues; see (\ref
{TQ6v1}), (\ref{TQ6v2}), (\ref{Qfus}), (\ref{TNQ}) and (\ref{fusTN}).

\end{document}